\documentclass[aps,twocolumn,prx]{revtex4-1}
\usepackage{graphicx,amsmath,amssymb,amsfonts,bm, color}

\usepackage{xcolor,hyperref}
\hypersetup{
   colorlinks,
   linkcolor={blue!50!black},
   citecolor={blue!50!black},
   urlcolor={blue!80!black}
}

\usepackage{enumitem}
\usepackage[normalem]{ulem}

\renewcommand{\=}{\!=\!}
\newcommand{\1}{^{\mbox{\tiny (1)}}}

\newcommand{\dbar}{{\,\mathchar'26\mkern-12mu d}}
\DeclareMathAlphabet{\mathitbf}{OML}{cmm}{b}{it}
\newcommand{\uv}{\mathitbf u}
\newcommand{\rv}{\mathitbf r}
\newcommand{\nv}{\mathitbf n}

\newcommand{\calBold}[1]{\mbox{\boldmath${\cal #1}$}}

\begin{document}

\title{Wave attenuation in glasses: Rayleigh and generalized-Rayleigh scattering scaling}
\author{Avraham Moriel$^1$}
\author{Geert Kapteijns$^2$}
\author{Corrado Rainone$^2$}
\author{Jacques Zylberg$^1$}
\author{Edan Lerner$^2$}
\thanks{e.lerner@uva.nl}
\author{Eran Bouchbinder$^1$}
\thanks{eran.bouchbinder@weizmann.ac.il}
\affiliation{$^1$ Chemical and Biological Physics Department, Weizmann Institute of Science, Rehovot 7610001, Israel\\
$^2$ Institute for Theoretical Physics, University of Amsterdam, Science Park 904, 1098 XH Amsterdam, The Netherlands}

\begin{abstract}
The attenuation of long-wavelength phonons (waves) by glassy disorder plays a central role in various glass anomalies, yet it is neither fully characterized, nor fully understood. Of particular importance is the scaling of the attenuation rate $\Gamma(k)$ with small wavenumbers $k\!\to\!0$ in the thermodynamic limit of macroscopic glasses. Here we use a combination of theory and extensive computer simulations to show that the macroscopic low-frequency behavior emerges at intermediate frequencies in finite-size glasses, above a recently identified crossover wavenumber $k_\dagger$, where phonons are no longer quantized into bands. For $k\!<\!k_\dagger$, finite-size effects dominate $\Gamma(k)$, which is quantitatively described by a theory of disordered phonon bands. For $k\!>\!k_\dagger$, we find that $\Gamma(k)$ is affected by the number of quasilocalized nonphononic excitations, a generic signature of glasses that feature a universal density of states. In particular, we show that in a frequency range in which this number is small, $\Gamma(k)$ follows a Rayleigh scattering scaling $\sim\!k^{\dbar+1}$ ($\dbar$ is the spatial dimension), and that in a frequency range in which this number is sufficiently large, the recently observed generalized-Rayleigh scaling of the form $\sim\!k^{\dbar+1}\log\!{(k_0/k)}$ emerges ($k_0\!>k_\dagger$ is a characteristic wavenumber). Our results suggest that macroscopic glasses --- and, in particular, glasses generated by conventional laboratory quenches that are known to strongly suppress quasilocalized nonphononic excitations --- exhibit Rayleigh scaling at the lowest wavenumbers $k$ and a crossover to generalized-Rayleigh scaling at higher $k$. Some supporting experimental evidence from recent literature is presented.
\end{abstract}

\maketitle

\section{Background and motivation}
\label{sec:intro}

Glasses feature distinct properties compared to their crystalline counterparts, where the differences are commonly attributed to the lack of long-range order in their characteristic disordered structures. Such differences manifest themselves in a broad range of physical properties, including those related to energy/heat transport. A crucial step in developing predictive theories of such transport properties is understanding how long-wavelength phonons (waves) --- the energy carriers --- are attenuated in glasses, and in particular elucidating the dependence of the phononic attenuation rate $\Gamma(k)$ on the wavenumber $k$ in the long-wavelength limit. In perfect crystals, phonons are expected to propagate indefinitely without attenuation. In this case, finite energy transport emerges due to anharmonicity, e.g.~phonon-phonon interactions~\cite{ashcroft_mermin}. In glasses, structural disorder can lead to the attenuation (scattering) of phonons even in the purely harmonic regime, which is expected to be realistic for low temperatures.

Consequently, the attenuation of phonons in glasses is intimately linked to the nature of glassy disorder. For a long time, it has been assumed --- with some inconclusive experimental support~\cite{experiments_300K_vSiO2,experiments_1620K_vSiO2,experiments_570K_dSiO2} --- that the phononic attenuation rate in glasses follows Rayleigh scattering scaling~\cite{Rayleigh}, $\Gamma(k)\!\sim\!k^{\dbar+1}$, in the long wavelength limit $k\!\to\!0$ ($\dbar$ is the spatial dimension). In this physical picture, glassy disorder plays a similar role to that of small, randomly distributed impurities that are responsible for wave scattering as originally described by Rayleigh~\cite{Rayleigh}. That is, at least as far as the attenuation of long-wavelength phonons is concerned, glassy disorder in this picture is basically equivalent to generic randomness without distinct features~\cite{ganter2010rayleigh,schirmacher2011comments,Maurer2004,mw_EM_epl,eric_boson_peak_emt}. On the other hand, recent developments in studying computer glass-formers have revealed distinct properties of glassy disorder and their manifestations on the physics of glasses.

An attenuation rate that follows a generalized-Rayleigh scaling of the form $\Gamma(k)\!\sim\!k^{\dbar+1}\log\!{(k_0/k)}$ ($k_0$ is some characteristic wavenumber) has been recently observed in computer glass simulations~\cite{lemaitre_tanaka_2016}. The logarithmic enhancement of wave attenuation, relative to Rayleigh scaling, has been therein attributed to long-range spatial correlations in the local elastic constants of glasses~\cite{lemaitre_tanaka_2016}. Subsequent work on jammed harmonic sphere packings reproduced the logarithmic enhancement, but also demonstrated that the removal of frustration-induced internal-stresses leads to its disappearance (i.e.~Rayleigh scaling is recovered)~\cite{Ikeda_2018}. Roughly at the same time, it was shown that frustration-induced internal-stresses, which are a generic property of glasses~\cite{shlomo}, are responsible for the existence of low-frequency quasilocalized nonphononic excitations in glasses~\cite{inst_note, ikeda_pnas}.

These nonphononic excitations have been hypothesized to exist for a long time~\cite{soft_potential_model_1991,Gurevich2003}, also in relation to the low-temperature anomalies mentioned above, but only recently their statistical and spatial properties have been directly revealed using computer glass simulations. It has been shown that the nonphononic excitations follow a universal density of states (DOS) $D_{\rm G}(\omega)\!\sim\!\omega^4$ ($\omega$ is the vibrational frequency), which is different from Debye's phononic DOS $D_{\rm D}(\omega)\!\sim\!\omega^{\dbar-1}$, and that unlike low-frequency phonons (which are spatially extended) they are quasilocalized in space (featuring a disordered core decorated with a power-law decay)~\cite{modes_prl_2016,modes_prl_2018}. Finally, recent computational advances~\cite{itamar_swap,berthier_prx} allow to obtain glassy states that are comparable to or even more stable than laboratory ones (while conventional computer glasses are generated using cooling rates that are orders of magnitude larger than laboratory ones). It has been shown that the universal DOS $D_{\rm G}(\omega)\!\sim\!\omega^4$ of nonphononic excitations remains valid for such deeply supercooled, experimentally-relevant glasses, but that their number (quantified by the prefactor in the DOS) is severely reduced~\cite{LB_modes_2019}.

Despite this progress, the possible implications of the existence and number of nonphononic excitations on glassy transport, and on wave attenuation in particular, have not be elucidated. Moreover, it remains unclear whether macroscopic glasses, generated by realistic laboratory protocols, feature Rayleigh or generalized-Rayleigh scaling in the long-wavelength limit. In this paper we use a combination of theory and extensive computer simulations to address the following fundamental, intrinsically-related questions: (i) What is the scaling of $\Gamma(k)$ in macroscopic glasses generated through conventional laboratory protocols? Addressing this question using computer glasses requires understanding finite size effects that should be systematically eliminated to expose the macroscopic limit. (ii) What is the physical origin of the generalized-Rayleigh scaling, when it is observed?

\section{Theoretical considerations: from finite size effects to Rayleigh and generalized-Rayleigh scaling}
\label{sec:theory}

To set the stage for addressing these questions, we discuss in this section the mathematical formulation that allows calculating the attenuation rate $\Gamma(k)$ and several theoretical predictions regarding $\Gamma(k)$.

{\em Mathematical formulation ---} We are interested in the attenuation rate of long-wavelength phononic excitations (plane-waves) that are introduced at time $t\=0$ as perturbations to metastable equilibrium glass states (inherent structures) characterized by particle positions ${\bm r}_i$, and take the form
\begin{equation}
\dot{\bm u}_i(t\=0) = {\bm a}_p\,\sin({\bm k}\!\cdot\!{\bm r}_i) \ ,
\label{eq:phonon}
\end{equation}
where $\dot{\bm u}_i$ are the particle velocities (hence ${\bm u}_i$ are the displacements), expressed in terms of the relevant natural units, as explained below. Here ${\bm a}_p$ is a unit polarization vector ($|{\bm a}_p|\=1$), $p\={\rm T}$ corresponds to transverse phonons (shear plane-waves) ${\bm a}_{\rm T}\!\cdot\!{\bm k}\!=\!0$, $p\={\rm L}$ corresponds to longitudinal phonons (dilatational plane-waves/sound) ${\bm a}_{\rm L}\!\cdot\!{\bm k}\=k$, and $k\=|{\bm k}|$ is the wavenumber corresponding to the wavevector ${\bm k}$. The wavevector is given by ${\bm k}\=(2\pi/L)\sum_{j=1}^\dbar n_j {\bm e}_j$, where $L$ is the linear size of the glass (hence its volume is $L^\dbar$), $j$ represents a Cartesian direction, ${{\bm e}_j}$ are orthonormal Cartesian vectors and ${n_j}$ are integers that are solutions to the sum of squares problem in $\dbar$-dimensions $\sum_{j=1}^\dbar n_j^2\=q$, where $q$ is an integer. The number of solutions to this problem for a given $q$ will be denoted hereafter by $n_q$, where we are mainly interested in relatively small $q$ values that correspond to long-wavelength excitations, $k\!\to\!0$. The glass-forming models and preparation protocols used in this work are presented in Appendix~\ref{sec:numerics_appendix}.

The only information about the glass that is incorporated into Eq.~\eqref{eq:phonon}, except for the particle positions ${\bm r}_i$, is its linear size $L$. As such, Eq.~\eqref{eq:phonon} represents long-wavelength phononic excitations that correspond to a perfectly ordered system, which are not eigenmodes of the glass Hessian ${\calBold M}$; the eigenmodes are determined by ${\calBold M}\cdot{{\bm \psi}_m}\=\omega^2_m {\bm \psi}_m$, where ${\bm \psi}_m$ is an eigenmode with an eigenvalue $\omega^2_m$, and $\omega_m$ is a vibrational frequency (assuming unity masses). The difference between the pure phononic excitation of  Eq.~\eqref{eq:phonon} and the set of the glass eigenmodes is at the heart of wave scattering in glasses.

To see this, we follow \cite{lemaitre_tanaka_2016} and define the $\dbar N$-dimensional velocity vector $\dot{\bm u}(t)$, composed of the velocity vectors $\dot{\bm u}_i(t)$ of all of the
particles that satisfy for $t\!>\!0$ the harmonic dynamics $\ddot{\bm u}_i(t)\=-\calBold{M}_{ij}\!\cdot\!{\bm u}_j(t)+\dot{\bm u}_i(t\=0)\,\delta(t)$, where here and in what follows repeated indices are summed over. The attenuation rate is probed through the time-evolution of the velocity autocorrelation function \cite{lemaitre_tanaka_2016}
\begin{equation}
\label{eq:corr}
  C(t) = \frac{\dot{\bm u}_i(t)\!\cdot\!\dot{\bm u}_i(t\=0)}{\dot{\bm u}_i(t\=0)\!\cdot\!\dot{\bm u}_i(t\=0)}=\sum_m \xi_m^2 \cos(\omega_m t) \ ,
\end{equation}
with
\begin{equation}
\label{eq:projections}
\xi_m = \frac{{\bm \psi}_{m,i}\!\cdot\!\dot{\bm u}_i(t\=0)}{\sqrt{\dot{\bm u}_i(t\=0)\!\cdot\!\dot{\bm u}_i(t\=0)}} \ ,
\end{equation}
where ${\bm \psi}_{m,i}$ is the $\dbar$-dimensional (Cartesian) vector component of ${\bm \psi}_{m}$ at the $i$th particle.

Equations \eqref{eq:corr}-\eqref{eq:projections} clearly show that wave scattering in glasses is of structural, not dynamical, origin; and that it is fully encapsulated in the normalized projections $\xi_m$ of the initial phononic excitation $\dot{\bm u}_i(t\=0)$ on the eigenmodes ${\bm \psi}_m$ of the glass Hessian. As $\Gamma(k)$ quantifies the characteristic decay rate of $C(t)$ in Eq.~\eqref{eq:corr}, it is intrinsically related to the spectral width over which the projections $\xi_m$ are sizable for a given $k$, as will be clarified below. Finally, $C(t)$ --- measured in simulations by averaging over independent glassy samples --- is commonly described by the damped oscillator model of the form
\begin{equation}
\label{eq:damped_oscillator}
C(t)\simeq\exp\left(-\tfrac{1}{2}\Gamma_{\rm T,L}(k)t\right)\cos\left(\Omega_{\rm T,L}(k)t \right) \ ,
\end{equation}
for both transverse and longitudinal phononic excitations. Equation~\eqref{eq:damped_oscillator} allows (if the damped oscillator model is valid, see discussion below) to extract the attenuation rate $\Gamma_{\rm T,L}(k)$ and wave speed $c_{\rm T,L}(k)\!\equiv\!\Omega_{\rm T,L}(k)/k$. In the remainder of the paper we will focus on the attenuation rate $\Gamma(k)$ \cite{footnote2}.

{\em Finite-size scaling theory of $\Gamma(k)$ ---} Our major goal, as stated above, is to understand the scaling properties of $\Gamma(k)$ in macroscopic glasses in the small $k$ limit. Experiments, using measurement techniques such as high-resolution inelastic x-ray scattering (IXS)~\cite{Ruocco1996,ixs_review_arXiv}, are insightful, but also suffer from some limitations (see Sect.~\ref{sec:exp}). Moreover, controlling and manipulating glassy disorder is highly challenging from an experimental perspective. Here we address the posed questions using computer glass simulations, which offer powerful opportunities to explore the physics of glasses in general, and in particular in the context of glassy disorder and structures.

One obvious limitation of present-day computer glass simulations is that computationally feasible system sizes are significantly smaller than the sizes of laboratory glasses. In the present context, this limitation may make it difficult to probe the macroscopic low-frequency/wavenumber regime using computer glasses. More specifically, it is clear that the lowest frequency/wavenumber response of computer glasses is dominated by finite-size effects; consequently, we need to understand these finite-size effects in order to systematically eliminate them, allowing to cleanly probe the macroscopic response. As will be shown below, understanding finite-size effects also provides basic insight into the general scattering problem.

The finite-size scaling theory of $\Gamma(k)$ has been developed very recently in~\cite{phonon_widths}. It starts with the observation that phonons of low-frequency $\omega$ in finite-size systems are quantized into discrete bands with a degeneracy $n_q(\omega)$, defined above through solutions of the integer sum of squares problem. The basic idea is that the presence of disorder lifts the degeneracy of the bands such that they are composed of disordered phonons of close, but not identical, frequencies. Consequently, the lowest frequency phononic bands feature finite spectral widths $\Delta\omega(\omega)$ that are smaller than the gaps between adjacent bands.

It is therefore clear that a pure phononic excitation that belongs to one of these finite width bands features sizable projections $\xi_m$ on disordered phonons within the band, but negligible projections on disordered phonons outside of the band. Hence, we expect $\Gamma(\omega)\!\sim\!\Delta\omega(\omega)$. Very recently, the spectral widths $\Delta\omega(\omega)$ have been theoretically derived using degenerate perturbation theory, random matrix theory and simple statistical considerations~\cite{phonon_widths}. The theory, which is described in great detail in~\cite{phonon_widths}, yields
\begin{equation}
\label{eq:finite-size}
\Gamma(\omega) \sim  \Delta\omega(\omega) \sim \frac{\omega\sqrt{n_q(\omega)}}{\sqrt{N}}  \ ,
\end{equation}
where $N\!\sim\!L^\dbar$ is the number of particles in the system. The explicit appearance of $N$, and of the degeneracy level $n_q(\omega)$ of discrete phononic bands of index $q$, make the finite-size nature of this result evident. The theoretical prediction in Eq.~\eqref{eq:finite-size} has been verified for disordered lattices, but not yet for glasses. We extensively test this prediction for glasses below and use it to eliminate finite-size effects, thus exposing the macroscopic limit. Finally, note that as the disordered phononic modes in the finite-size regime satisfy wave dispersion $\omega\!\propto\! k$ (where the proportionality factor is either the transverse or the longitudinal wave-speed, with transverse phonons populating the lowest frequencies), Eq.~\eqref{eq:finite-size} is equally valid for $\Gamma(k)$ once $\omega$ is everywhere replaced by $k$.

{\em Derivation of a relation between $\Gamma(k)$ and the phononic DOS $D_{\rm D}(\omega)$, and of Rayleigh scattering scaling ---} Equation~\eqref{eq:finite-size} predicts that the width of disordered phononic bands increases with $\omega$ for a fixed $N$. On the other hand, the spectral gap between adjacent phononic bands decreases with increasing $\omega$. Consequently, there exists a characteristic frequency $\omega_\dagger$ for which the band width $\Delta\omega$ becomes comparable to the gap between adjacent bands and above which discrete phononic bands cease to be well-defined. The crossover frequency $\omega_\dagger$ has been shown to take the form~\cite{phonon_widths}
\begin{equation}
\label{eq:omega_dagger}
\omega_\dagger(N) \sim L^{-2/(\dbar + 2)} \sim N^{-2/\dbar(\dbar + 2)} \sim k_\dagger(N) \ .
\end{equation}

It is clear that the finite-size scaling theory of Eq.~\eqref{eq:finite-size} is valid for $\omega\!<\!\omega_\dagger$. What happens for $\omega\!>\!\omega_\dagger$? As explained above, discrete phononic bands do not exist anymore for $\omega\!>\!\omega_\dagger$, rather disordered phonons form a continuum that is expected to follow Debye's DOS $D_{\rm D}(\omega)$. In order to understand wave attenuation in this regime, we are interested in the spectral width $\Delta\omega(\omega)$ of bands comprised of disordered phonons, on which a pure phononic excitation of frequency $\omega$ has sizable projections $\xi_m$. The number of such disordered phonons is simply $N D_{\rm D}(\omega)\Delta\omega(\omega)$. While these disordered phonons do not necessarily emerge from strictly degenerate phonons in the corresponding ordered problem (as in the case of discrete bands), they are clearly nearly degenerate and hence their number plays the role of $n_q(\omega)$ of Eq.~\eqref{eq:finite-size}.

With this identification, we can substitute $n_q(\omega)\!\sim\!N D_{\rm D}(\omega)\Delta\omega(\omega)$ in Eq.~\eqref{eq:finite-size} and treat it as a self-consistency equation for the spectral width $\Delta\omega(\omega)$ in the $\omega\!>\!\omega_\dagger$ regime. Solving for $\Delta\omega(\omega)$, and using $\Gamma(\omega)\!\sim\!\Delta\omega(\omega)$, we readily obtain
\begin{equation}
\label{eq:Gamma_DOS}
\Gamma(\omega) \sim \omega^2 D_{\rm D}(\omega)\qquad\hbox{for}\qquad  \omega>\omega_\dagger\ ,
\end{equation}
which is a fundamental relation between the wave attenuation rate $\Gamma(\omega)$ and Debye's phononic DOS $D_{\rm D}(\omega)$. This relation is known as the ``Rayleigh-Klemens law"~\cite{Klemens1951}, and has been obtained in the past using quite different considerations and mathematical tools~\cite{mw_EM_epl,eric_boson_peak_emt,john1990localization, Maurer2004}. We believe that the derivation just presented, based on the analogy with the finite-size scaling theory in Eq.~\eqref{eq:finite-size}, provides basic and transparent understanding of the physical content of Eq.~\eqref{eq:Gamma_DOS}.

An immediate corollary of Eq.~\eqref{eq:Gamma_DOS}, simply by recalling that $D_{\rm D}(\omega)\!\sim\!\omega^{\dbar-1}$ and $\omega\!\sim\!k$, reads $\Gamma(k)\!\sim\! k^{\dbar+1}$ for $k\!>\! k_\dagger$, which is nothing but the well-known Rayleigh scattering scaling~\cite{Rayleigh}. We summarize this discussion with the prediction
\begin{equation}
\label{eq:scattering_summary}
\Gamma(k) \sim \left\{ \begin{array}{cc}k\sqrt{n_q(k)}/\sqrt{N}\,,\vspace{0.3cm}&k<k_\dagger\sim N^{-2/\dbar(\dbar + 2)}\\k^{\dbar+1}\,,&k>k_\dagger\sim N^{-2/\dbar(\dbar + 2)}\end{array}\right.\,,
\end{equation}
which we will verify using numerical simulations in the subsequent sections.

{\em Quasilocalized nonphononic excitations ---} The theoretical considerations culminating in Eq.~\eqref{eq:scattering_summary} suggest that had the elementary low-frequency excitations in glasses been exclusively comprised of disordered phonons, we would expect the wave attenuation rate of macroscopic glasses $\Gamma(k)$, in the small wavenumber limit, to exhibit Rayleigh scattering scaling. Moreover, had it been the case, we expect finite-size computer glasses to reveal this scaling in the range $k\!>\!k_\dagger$, once the finite-size effects in the range $k\!<\!k_\dagger$ are properly identified and eliminated. However, as was already stated in Sect.~\ref{sec:intro}, it is now well-established that glasses feature in addition to disordered phonons also low-frequency nonphononic excitations, which are characterized by quasilocalized spatial structures and follow a universal non-Debye DOS $D_{\rm G}(\omega)\!\sim\!\omega^4$.

These nonphononic modes --- sometimes termed glassy modes --- are known to hybridize with disordered phonons when they both coexist in the same frequency range \cite{SciPost2016,phonon_widths}. As such, and in light of the basic understanding that $\Gamma(k)$ is encapsulated in the projections $\xi_m$ of a pure phononic excitation of wavenumber $k$ on the eigenmodes of the glass Hessian, one expects glassy modes to affect wave attenuation in glasses. In fact, preliminary evidence in the literature indicates that glassy modes might be related to the generalized-Rayleigh scaling $\Gamma(k)\!\sim\!k^{\dbar+1}\log\!{(k_0/k)}$ (where $k_0$ is some characteristic wavenumber) recently observed in~\cite{lemaitre_tanaka_2016}; this connection may be implied by the observation of~\cite{Ikeda_2018} that the generalized-Rayleigh scaling disappears in favor of Rayleigh scaling once frustration-induced internal-stresses, which are known to be responsible for low-frequency nonphononic excitations in glasses~\cite{inst_note, ikeda_pnas}, are removed from jammed packings of harmonic spheres~\cite{footnote}.

In the subsequent sections we test the theoretical picture discussed here and the outstanding questions that accompany it. We will first test the finite-size scaling theory of $\Gamma(k)$ in the range $k\!<\!k_\dagger$ and use it to identify the macroscopic regime in the range $k\!>\!k_\dagger$. We will then explore the scaling of $\Gamma(k)$ in the range $k\!>\!k_\dagger$, focusing on the existence of Rayleigh and/or generalized-Rayleigh scaling regimes and on their possible relation to low-frequency nonphononic excitations. We note that macroscopic glasses are also known to feature a higher-$k$ scaling regime $\Gamma(k)\!\sim\!k^2$ above a characteristic wavenumber $k_*$, which is related to the Boson peak~\cite{experiments_1620K_vSiO2,Ruta_JNCS_2011,eric_boson_peak_emt}. This scaling is consistent with Eq.~(\ref{eq:Gamma_DOS}) for a flat $D(\omega)$ as observed in the vicinity of the Boson peak frequency in studies of the unjamming transition \cite{ohern2003,liu_review,van_hecke_review}. While some of our data sets will demonstrate this quadratic scaling regime as well, our major focus is on the range $k_\dagger\!<\!k\!<\!k_*$.

In what follows, simulational observables are made dimensionless: lengths are reported in terms of $a_0\!\equiv\! (V/N)^{1/\dbar}$ where $V,N,\dbar$ denote the volume, number of particles, and spatial dimension, respectively; times (rates) are reported in units of $a_0/c_{\rm T}$ ($c_{\rm T}/a_0$), where $c_{\rm T}\!\equiv\!\sqrt{G/\rho}$ denotes the shear wave speed, $G$ denotes the athermal shear modulus and $\rho$ denotes the mass density.

\section{Eliminating finite-size effects reveals the macroscopic low-frequency scaling regime}
\label{sec:2dipl}

As explained in the previous section, wave attenuation rates in finite-size glasses are expected to be affected by the quantization of disordered phonons into discrete bands at frequencies $\omega\!<\!\omega_\dagger$ (cf.~Eq.~(\ref{eq:finite-size})), or, alternatively, at wavenumbers $k\!<\!k_\dagger$ (cf.~Eq.~(\ref{eq:scattering_summary})). This expectation is verified in our simulational data presented in Fig.~\ref{fig:Fig1} below, generated using a generic computer glass forming model, referred to hereafter as the 2DIPL model. A comprehensive description of this model, and of other employed models and measurement methods, is provided in Appendix~\ref{sec:numerics_appendix}; here we briefly note that 2DIPL glassy samples were quenched from the equilibrium liquid phase using a constant and relatively large quench rate. This results in glasses that possess an abundance of soft quasilocalized modes, as discussed previously (e.g.~in \cite{modes_prl_2018}) and in what follows.
\begin{figure*}[ht!]
\centering
\includegraphics[width=1.9\columnwidth]{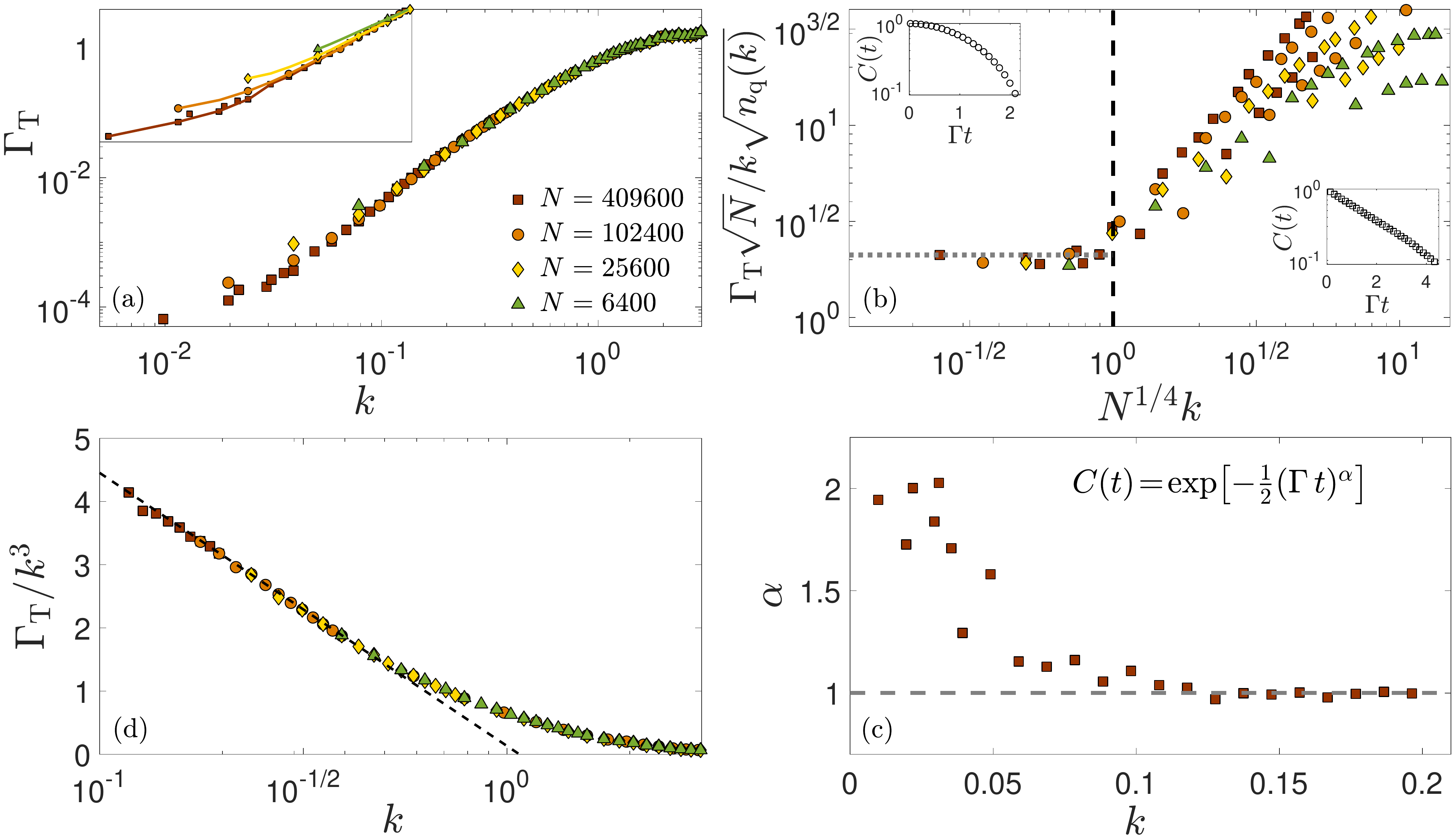}
\caption{Finite-size effects in wave attenuation rates in computer glasses and their systematic elimination. Results are shown for the 2DIPL model (see Appendix~\ref{sec:numerics_appendix} for model details and methods). (a) The attenuation rate $\Gamma_{\rm T}$ of transverse waves vs.~wavenumber $k$ for various system sizes (see legend). The inset zooms in on the meandering away from the $N$-independent curve at low $k$. (b) The same data as in panel (a), but rescaled following the finite-size predictions for $\Gamma(k,n_q,N)$ and $k_\dagger(N)$ given in Eq.~(\ref{eq:scattering_summary}). The data collapse (see horizontal dashed line) at low $k$ allows to estimate the crossover wavenumber $k_\dagger(N)$ (vertical dashed line). The insets show the accompanying crossover in the functional form of the envelopes $C(t)$ of the velocity autocorrelation functions from exponential (bottom-right inset) to compressed-exponential (upper-left inset). (c) The exponent $\alpha(k)$ obtained by fitting the envelope of $C(t)$ to a compressed-exponential form (here $N\!=\!409600$) vs.~wavenumber $k$. A transition from $\alpha\!=\!1$ for $k\!>\! k_\dagger$ to $\alpha\!\approx2$ for $k\!<\! k_\dagger$ is observed. (d) The same data as in panel (a), once finite-size effects are eliminated, and $\Gamma_{\rm T}$ is rescaled by $k^3$. The dashed line corresponds to a generalized-Rayleigh scaling of attenuation rates, see text for additional discussion.}\label{fig:Fig1}
\end{figure*}

In Fig.~\ref{fig:Fig1}a we show the bare attenuation rates of transverse waves $\Gamma_{\rm T}(k)$ vs.~wavenumber $k$, measured in 2DIPL glassy samples for various system sizes, as indicated in the legend. A regime of wavenumbers in which the attenuation rates are \emph{independent} of system size is clearly identified. This independence holds down to an $N$-dependent crossover wavenumber $\omega_\dagger(N)$, as expected from our theoretical considerations. Below this crossover wavenumber, attenuation rates begin to meander away from the $N$-independent curve (see inset of Fig.~\ref{fig:Fig1}a). Similar finite-size-induced deviations were discussed in~\cite{lemaitre_tanaka_2016}.

According to our theoretical considerations (cf.~Eq.~(\ref{eq:omega_dagger}) and preceding discussions), the crossover wavenumber $k_\dagger$, below which finite-size effects on attenuation rates should be observed, decreases with system size as $N^{-1/4}$ in 2D glasses. Below this crossover, attenuation rates are expected to follow the finite-size scaling given by Eq.~(\ref{eq:scattering_summary}), i.e.~$\Gamma\!\sim\! k\sqrt{n_q(k)}/\sqrt{N}$. To test these predictions, we plot in Fig.~\ref{fig:Fig1}b the attenuation rates $\Gamma(k)$ rescaled by $k\sqrt{n_q(k)}/\sqrt{N}$, against the rescaled wavenumber $N^{1/4}k$. The data collapse to a constant curve below the crossover wavenumber $k_\dagger\!\sim\! N^{-1/4}$ (vertical dashed line) validates both the prediction for the scaling of the crossover wavenumber $k_\dagger(N)$ given by Eq.~(\ref{eq:omega_dagger}), and the prediction for the wavenumber $k$, phonon-band degeneracy $n_q$, and system size $N$ dependencies of attenuation rates for $k\!<\!k_\dagger$, given by Eq.~(\ref{eq:scattering_summary}).

The change in the scaling behaviour of attenuation rates at $k\!\!\!\!<\!\!\!\! k_\dagger$ is accompanied by a qualitative change in the \emph{functional form} of the envelope of the correlation function $C(t)$, as shown in Fig.~\ref{fig:Fig1}b (insets). While exponentially-decaying oscillations, consistent with Eq.~\eqref{eq:damped_oscillator}, are observed for $k\!\!>\!\! k_\dagger$, compressed-exponentially decaying oscillations are observed for~$k\!<\! k_\dagger$. To quantify this qualitative change in functional form, we fitted the envelope of the autocorrelation function $C(t)$ to a compressed-exponential $\exp\left[-\tfrac{1}{2}(\Gamma(k)t)^{\alpha(k)}\right]$, where the exponent $\alpha(k)$ is shown in Fig.~\ref{fig:Fig1}c. It is observed that while the conventional assumption of exponential decay (cf.~Eq.~\eqref{eq:damped_oscillator}) is valid above $k_\dagger$, it breaks down below it, where $\alpha$ is larger and appears to approach $2$ at the lowest $k$'s. Consequently, care should be taken when extracting attenuation rates in the finite-size regime $k\!<\! k_\dagger$. We expand on this point and on our fitting procedure in Appendix~\ref{sec:fitting_procedure}.

In order to understand the crossover from the exponential decay of oscillations at $k\!>\!k_\dagger$ to compressed exponential decay of oscillations for $k\!<\! k_\dagger$, one should reexamine Eqs.~\eqref{eq:corr}-\eqref{eq:projections}; from these relations it is clear that $C(t)$ (which becomes an even function of time by replacing $t\!\to\!|t|$) is approximately given by the Fourier transform of the squared overlaps $\xi_m^2$ weighed by the DOS, which in turn depends on the wavenumber $k$ (or, alternatively, on the frequency $\omega$). In the finite-size regime, in which phonon bands are well-isolated from each other, the DOS-weighed $\xi_m^2$ is sharply peaked (as shown e.g.~in \cite{lemaitre_tanaka_2016}), with a highly confined support on the frequency axis. This leads to the expectation that the Fourier transform of the DOS-weighed $\xi_m^2$ should approximately feature a Gaussian envelope (as holds e.g.~for the Fourier transform of a Gaussian), as we indeed observe. Once the support of $\xi_m^2$ becomes unconfined, as occurs for $k\!>\! k_\dagger$, $C(t)$ features exponentially-decaying oscillations. Finally, note that in recent work~\cite{Grzegorz_scattering_arXiv2} it has been claimed that at very short times $C(t)$ features an exponential decay in the finite-size regime $k\!<\! k_\dagger$; we comment on this claim in Appendix~\ref{sec:networks_appendix}.

Identifying the finite-size effects on the attenuation rates, made apparent by the presentation scheme of Fig.~\ref{fig:Fig1}b, provides us with a solid handle on systematically eliminating them from simulation measurements. Consequently, we follow below the very same procedure to identify the crossover wavenumber $k_\dagger$ for each data set. In order to properly probe the scaling behaviour of macroscopic glasses, we shall only consider attenuation rates associated with wavenumbers $k$ sufficiently \emph{larger} than the identified crossover $k_\dagger$. Applying this scheme to our 2DIPL wave attenuation rates data, we replot  $\Gamma(k)$ in Fig.~\ref{fig:Fig1}d, this time rescaled by $k^3$ (Rayleigh scaling prediction for $\dbar\=2$), keeping only data points that fall \emph{above} the crossover wavenumber $k_\dagger$. We find that $\Gamma(k)$ in this system follows a generalized-Rayleigh law $\Gamma(k)\!\sim\!k^{\dbar+1}\log\!{(k_0/k)}$ with $k_0\!\approx\!1.07$. This scaling, first observed by Gelin et al.~\cite{lemaitre_tanaka_2016} and subsequently in~\cite{Ikeda_2018}, appears to persist down to the lowest wavenumber accessible above the finite-size scaling wavenumber regime $k\!>\! k_\dagger$. We note that $k_0$ close to unity corresponds to a length of about $6$ interparticle distances, very similar to the size of the core of quasilocalized modes observed in computer glass models \cite{modes_prl_2016,SciPost2016,modes_prl_2018}. Indeed, we next establish that the abundance of quasilocalized modes of frequencies comparable to $\omega_\dagger$ --- which depends in turn on glass stability \cite{protocol_prerc, cge_paper, LB_modes_2019} --- gives rise to the observed generalized-Rayleigh scaling of attenuation rates.

\section{The generalized-Rayleigh scaling depends on glass stability}
\label{sec:2d3dSWAP}

As mentioned in Sect.~\ref{sec:theory}, recent computer simulations by Mizuno and Ikeda~\cite{Ikeda_2018} have shown that wave attenuation rates measured in packings of harmonic spheres both in 2D and 3D follow Rayleigh scattering scaling $\Gamma\!\sim\! k^{\dbar+1}$ once the internal-stresses/forces between particles are artificially turned off \cite{footnote}, while the generalized-Rayleigh scaling $\Gamma\!\sim\! k^{\dbar+1}\log\!{(k_0/k)}$ is observed when these internal-stresses/forces are left intact. In other work \cite{inst_note, ikeda_pnas} it was shown that the very same procedure of turning off forces between particles suppresses the occurrence of soft quasilocalized modes, whose universal statistics have been recently established \cite{modes_prl_2016,cge_paper, modes_prl_2018}. These two observations lead together to the hypothesis that the emergence of generalized-Rayleigh scaling depends on the abundance of soft quasilocalized modes, and possibly on their characteristic localization length, both of which have been shown to depend, in turn, on glass stability \cite{protocol_prerc, cge_paper, LB_modes_2019}.

To test this hypothesis we seek to substantially increase the stability of our computer glasses, and by doing so to suppress the occurrence and spatial extent of soft quasilocalized modes in a physical manner. To this aim we employ the Swap Monte Carlo method and an optimized glass forming model~\cite{berthier_prx}, which was recently shown to be very efficient in 2D~\cite{two_dimensions_swap}, allowing for extremely deep supercooling. Details about our simulations and employed methods are provided in Appendix~\ref{sec:numerics_appendix}.
\begin{figure}[ht!]
  \centering
  \includegraphics[width=\columnwidth]{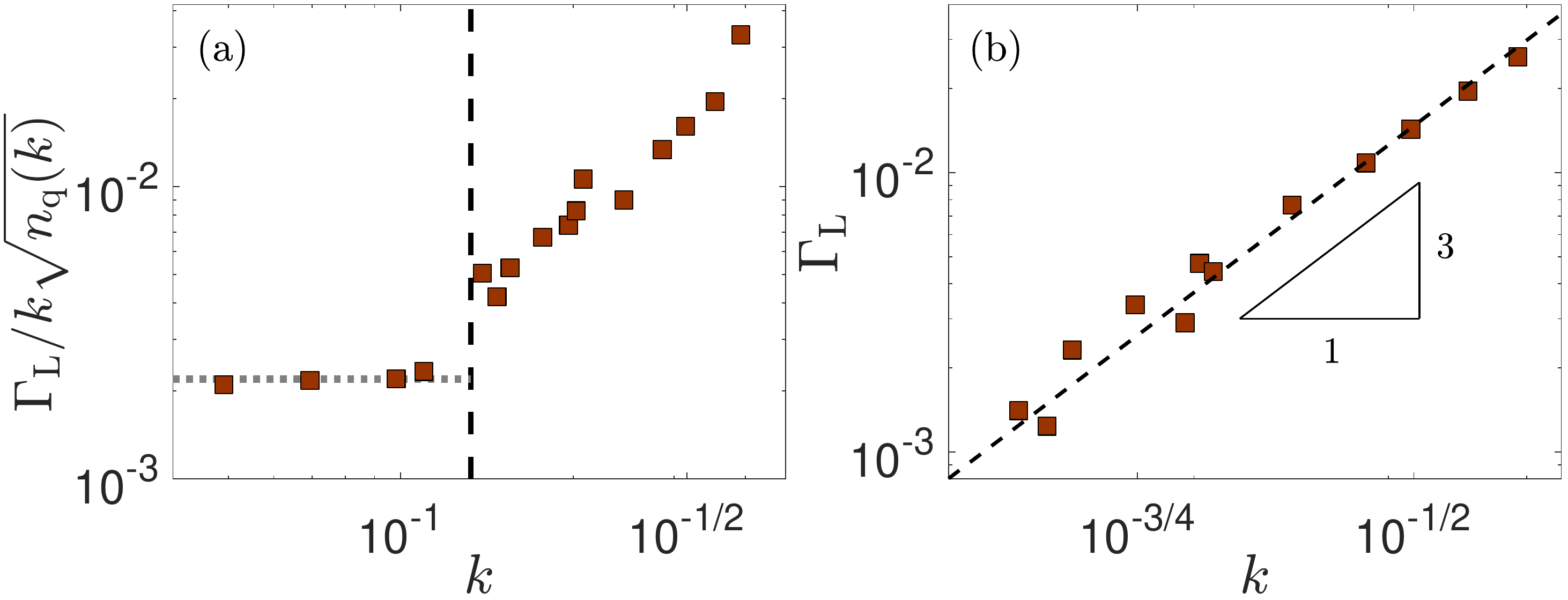}
  \caption{Sound wave attenuation rates in a stable 2D computer glass of $N\!=\!16384$. Results are shown for the 2DSWAP model (see Appendix~\ref{sec:numerics_appendix} for model details and methods). (a)~The attenuation rate $\Gamma_{\rm L}$ of longitudinal waves vs.~wavenumber $k$, rescaled according to the finite-size prediction for $\Gamma(k,n_q)$ given in Eq.~(\ref{eq:scattering_summary}). The predicted finite-size effect is verified (horizontal dashed line), along with an estimate of the accompanying crossover wavenumber $k_\dagger$ (vertical dashed line). (b) Rayleigh scattering scaling $\Gamma_{\rm L}\!\sim\! k^3$ is observed (dashed line, $1\!:\!3$ triangle) once finite-size effects are eliminated.}\label{fig:Fig2}
\end{figure}

In Fig.~\ref{fig:Fig2} we present the attenuation rates of longitudinal (sound) waves $\Gamma_{\rm L}(k)$ measured in very stable 2D computer glasses of $N\!=\!16384$ particles, referred to as the 2DSWAP model. Figure~\ref{fig:Fig2}a employs the same presentation scheme discussed in the preceding section, namely we plot $\Gamma_{\rm L}/k\sqrt{n_q}$ vs.~the wavenumber $k$ (here we do not vary $N$), in order to identify the crossover wavenumber $k_\dagger$. As expected, our data show that at the lowest wavenumbers $\Gamma_{\rm L}\!\sim\! k\sqrt{n_q(k)}$, apparently at odds with the results of~\cite{Grzegorz_scattering_arXiv}, where sound (longitudinal) wave attenuation rates in stable computer glasses in 3D appeared to follow Rayleigh scaling at the lowest accessible wavenumbers. This discrepancy stems from the differences in the location of the crossover frequency $\omega_\dagger$ relative to the lowest sound waves' frequencies: in Appendix~\ref{sec:soundwaves_appendix} we show that in our 2D stable glasses several discrete phonon bands that pertain to sound waves exist \emph{below} the crossover frequency $\omega_\dagger$, whereas in stable 3D glasses only a single sound wave band resides below the crossover frequency $\omega_\dagger$. We therefore suggest that if larger 3D stable glasses could have been generated, more sound waves would appear as discrete phonon bands in the vibrational density of states, resulting in sound wave attenuation rates that follow the finite-size scaling theory at the lowest accessible wavenumbers, as shown in Fig.~\ref{fig:Fig2} for 2D.

Having identified the crossover wavenumber $k_\dagger$, we next plot in Fig.~\ref{fig:Fig2}b the attenuation rates of longitudinal waves $\Gamma_{\rm L}(k)$ vs.~wavenumber $k$, keeping only data points that pertain to wavenumbers $k\!>\! k_\dagger$. We find that the enhanced stability of the studied model glasses results in a Rayleigh-like scaling $\sim\!k^3$ (in 2D) of attenuation rates of longitudinal waves, rather than the generalized-Rayleigh scaling as seen in less stable glasses in the preceding section. This observation supports the idea that the abundance of soft quasilocalized excitations gives rise to generalized-Rayleigh scaling of wave attenuation rates, and that Rayleigh scattering scaling is observed in their absence.

\begin{figure}[ht!]
  \centering
  \includegraphics[width=\columnwidth]{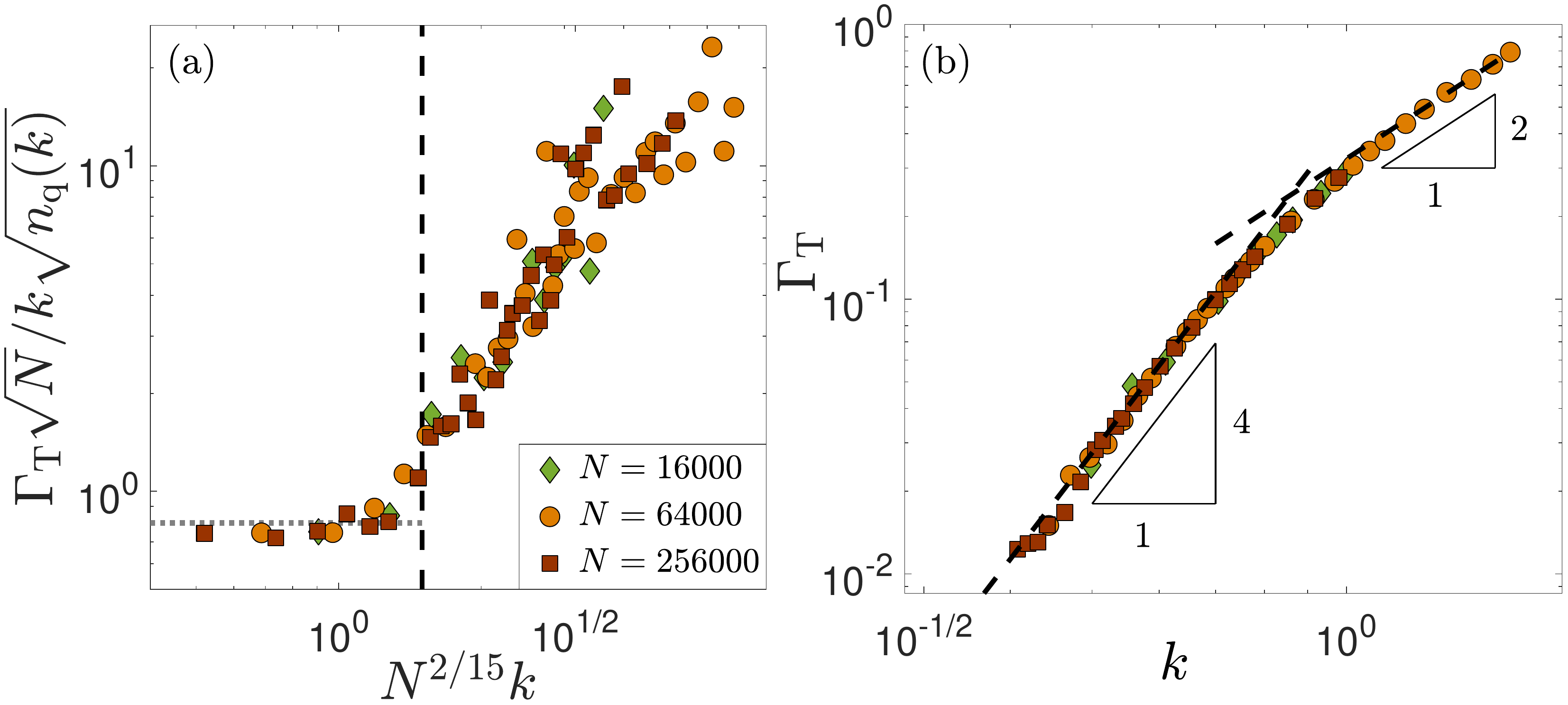}
  \caption{Transverse (shear) wave attenuation rates in stable 3D computer glasses. Results are shown for the 3DSWAP model (see Appendix~\ref{sec:numerics_appendix} for model details and methods). (a) The attenuation rate $\Gamma_{\rm T}$ of transverse waves vs.~wavenumber $k$, rescaled according to the finite-size predictions for $\Gamma(k,n_q,N)$ and $k_\dagger(N)$ given in Eq.~(\ref{eq:scattering_summary}). The predicted finite-size effect is verified (horizontal dashed line), along with an estimate of the accompanying crossover wavenumber $k_\dagger$ (vertical dashed line). (b) Rayleigh scattering scaling $\Gamma_{\rm T}\!\sim\! k^4$ is observed (dashed line, $1\!:\!4$ triangle) at low $k$ once finite-size effects are eliminated. A quadratic scaling regime $\Gamma_{\rm T}\!\sim\! k^2$ is observed above $k\!\approx\!1$ (dashed line, $1\!:\!2$ triangle).}\label{fig:Fig3}
\end{figure}

To further establish the generality of this conclusion, we also created an ensemble of stable glasses in 3D (see details in Appendix~\ref{sec:numerics_appendix}), referred to as the 3DSWAP system, and measured the attenuation rates $\Gamma_{\rm T}(k)$ of transverse waves. In Fig.~\ref{fig:Fig3} we employ the presentation scheme of Figs.~\ref{fig:Fig1}b and \ref{fig:Fig2}a for shear wave attenuation rates, to identify the onset of the finite size regime for each system size. The data collapse below the crossover wavenumber $\sim\! N^{-2/15}$ to a constant curve validates once again our theoretical predictions given by Eqs.~(\ref{eq:omega_dagger}) and (\ref{eq:scattering_summary}), this time in 3D.

In Fig.~\ref{fig:Fig3}b we plot the shear wave attenuation rates measured in the 3DSWAP system, after identifying and eliminating finite-size effects as described above. At the lowest wave numbers above $k_\dagger$ we find a very clean Rayleigh scattering scaling $\Gamma_{\rm T}\!\sim\! k^4$. A similar observation was recently made in~\cite{Grzegorz_scattering_arXiv}. These results reinforce our assertion that the abundance of soft quasilocalized modes gives rise to the generalized-Rayleigh scaling, and that, in their absence, Rayleigh scattering scaling is observed.

We note that our 3D data show the onset of a quadratic scaling of wave attenuation rates $\Gamma_{\rm T}\!\sim\! k^2$ above $k\!\approx\!1$, consistent with the Effective Medium Theory~\cite{eric_boson_peak_emt} and experimental observations~\cite{experiments_1620K_vSiO2,Ruta_JNCS_2011}. We further note, however, that wave attenuation rates are of the order of unity in this regime, indicating that the decay time of the velocity autocorrelation function becomes comparable to the wave oscillation time. In turn, this implies that the wavenumber window in which quadratic scaling is observed is quite limited in stable glasses.

\section{Numerical evidence for a crossover from Rayleigh to generalized-Rayleigh scaling}
\label{sec:crossover}

In the previous sections we established that if soft quasilocalized modes are too scarce, wave attenuation rates at low $k$ (but still above $k_\dagger$) follow Rayleigh scattering scaling $\Gamma\!\sim\! k^{\dbar+1}$, whereas generalized-Rayleigh scaling $\Gamma\!\sim\! k^{\dbar+1}\log\!{(k_0/k)}$ is observed in less stable glasses, in which soft quasilocalized modes are abundant. Recalling that the universal non-Debye DOS $D_{\rm G}(\omega)\!\sim\!\omega^4$ decays rather rapidly as $\omega\!\to\!0$, it is reasonable to expect that at frequencies corresponding to phonons/waves with very low wavenumbers, the abundance of soft quasilocalized modes would become so low that attenuation rates would cross over from the generalized-Rayleigh scaling to Rayleigh scattering scaling. In other words, based on the evidence presented in preceding sections we hypothesize that even in poorly-annealed glasses in which soft quasilocalized modes are abundant, an onset wavenumber $k_{\rm on}$, below which wave attenuation rates should cross over from $\Gamma\!\sim\! k^{\dbar+1}\log\!{(k_0/k)}$ at $k\!>\!k_{\rm on}$ to $\Gamma\!\sim\! k^{\dbar+1}$ at $k_\dagger(N)\!<\! k\!<\! k_{\rm on}$, should exist.

Testing this hypothesis requires creating glasses for which the crossover frequency $\omega_\dagger\!=\! c_{\rm T}k_\dagger$ is very low, achievable by increasing the system size, as evident from Eq.~(\ref{eq:omega_dagger}). While larger lengths and longer simulation runs are generally more feasible in 2D, we do not use 2D glasses in the present context. The reason is that the characteristic frequency of soft quasilocalized modes in 2D also decreases with system size, as shown in~\cite{cge_paper, modes_prl_2018}. This leads to a competition between soft quasilocalized modes' frequencies and the crossover frequency $\omega_\dagger$, both decreasing with system size, thus making it difficult to observe the hypothesized crossover in the scaling form of wave attenuation rates in 2D.

In 3D, the characteristic frequencies of soft quasilocalized modes is system-size independent, and we therefore opt to test the aforementioned crossover hypothesis in 3D. We employ a generic computer glass forming model --- referred to as the 3DIPL model --- and generate very large glassy samples of up to $4$M particles using a continuous, rapid quench from equilibrium liquid states (see Appendix~\ref{sec:numerics_appendix} for further details about the model and protocols). These glassy samples have been shown to feature an abundance of soft quasilocalized modes~\cite{phonon_widths, cge_paper, modes_prl_2018}.

\begin{figure}[ht!]
  \centering
  \includegraphics[width=0.85\columnwidth]{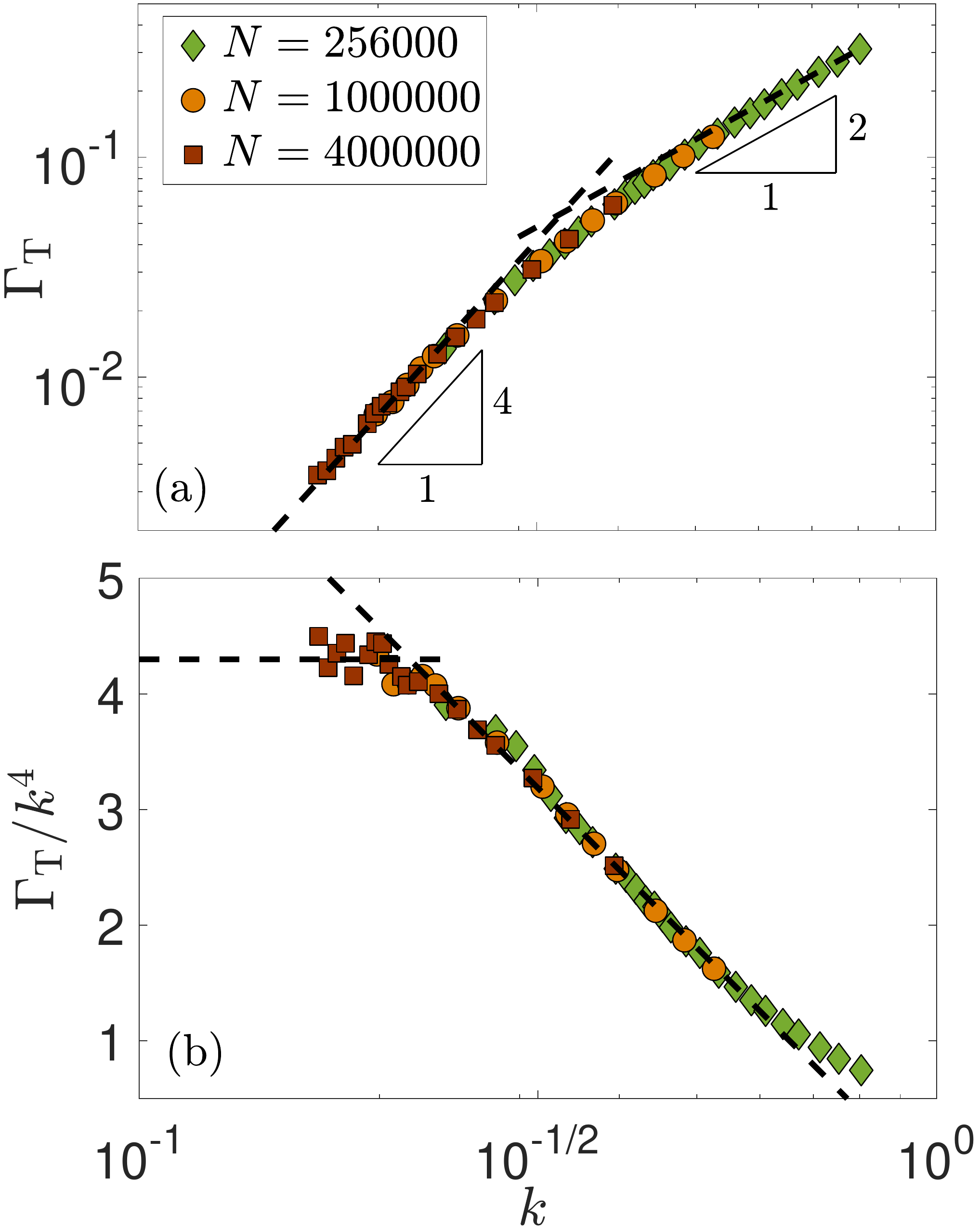}
  \caption{(a) Transverse (shear) wave attenuation rates $\Gamma_{\rm T}(k)$ measured in 3DIPL computer glasses of various sizes (see legend), plotted against wavenumber $k$. Data shown only for $k\!>\! k_\dagger$, as described in Sect.~\ref{sec:2dipl}. Rayleigh scattering scaling $\Gamma_{\rm T}\!\sim\! k^4$ is observed (dashed line, $1\!:\!4$ triangle) at low $k$ and a quadratic scaling regime $\Gamma_{\rm T}\!\sim\! k^2$ is observed at higher $k$ (dashed line, $1\!:\!2$ triangle). (b) Plotting $\Gamma(k)/k^4$ reveals that attenuation rates cross over from generalized-Rayleigh scaling at intermediate $k$ (tilted dashed line) to Rayleigh scaling at low $k$ (horizontal dashed line). The crossover, whose wavenumber $k_{\rm on}$ is roughly estimated by the intersection of the two dashed lines, occurs well within the regime in which the universal non-Debye DOS $D_{\rm G}(\omega)\!\sim\!\omega^4$ of soft quasilocalized modes is observed, as shown in Appendix~\ref{sec:DOS_appendix}.}\label{fig:Fig4}
\end{figure}

In Fig.~\ref{fig:Fig4} we present the shear wave attenuation rates as measured in the 3DIPL model. As was the case with previous data sets shown above and as explained in Sect.~\ref{sec:2dipl}, here too we identified and eliminated finite-size effects, showing only data for $k\!>\! k_\dagger(N)$. Figure~\ref{fig:Fig4}a shows the attenuation rates $\Gamma_{\rm T}(k)$ vs.~wavenumber $k$; $\Gamma_{\rm T}(k)$ appears to approach Rayleigh scattering at the lowest wavenumbers, and transitions at sufficiently high wavenumbers to the expected quadratic scaling $\Gamma_{\rm T}\!\sim\! k^2$, both marked by dashed lines.

To better assess the scaling form of attenuation rates at intermediate wavenumbers, we plot in Fig.~\ref{fig:Fig4}b the rescaled rates $\Gamma_{\rm T}(k)/k^4$ against $k$; this presentation reveals that at intermediate wavenumbers the generalized-Rayleigh scaling $\Gamma_{\rm T}\!\sim\! k^{\dbar+1}\log\!{(k_0/k)}$ emerges, consistent with previous observations~\cite{lemaitre_tanaka_2016,Ikeda_2018}. Our key result, however, is the crossover at low $k$ from generalized-Rayleigh scaling to Rayleigh scattering scaling $\Gamma\!\sim\! k^4$, which is hinted at by our data for systems of 1M particles (orange circles), but robustly observed for our largest systems of 4M particles (brown squares). This result strongly suggests that macroscopic glasses always exhibit Rayleigh scattering scaling at low wavenumber, \emph{regardless of glass stability}. This suggestion is further supported by experimental data from recent literature, discussed at length in Sect.~\ref{sec:exp} below. We stress that the crossover observed in Fig.~\ref{fig:Fig4}b occurs well within the regime in which the $D_{\rm G}(\omega)\!\sim\!\omega^4$ DOS is realized in our calculations, as shown in Appendix~\ref{sec:DOS_appendix}.

A similar crossover was recently announced by Mizuno and Ikeda~\cite{Ikeda_2018}. These authors performed measurements of wave attenuation rates in packings of particles interacting via one-sided harmonic potentials both in 2D and 3D, a well-studied model in the context of the unjamming transition~\cite{ohern2003,liu_review,van_hecke_review}. For the largest systems employed in~\cite{Ikeda_2018}, of over 4M particles in 3D, a clear finite-size quantization of the lowest frequency phonon bands is observed, see e.g.~Fig.~1A of~\cite{ikeda_pnas}. Despite this inevitable low wavenumber quantization, it was found in~\cite{Ikeda_2018} that Rayleigh scattering scaling appears to persist all the way down to the lowest accessible wavenumbers --- and thus necessarily for $k\!<\!k_\dagger$ --- where we expect the finite-size scaling theory Eq.~(\ref{eq:scattering_summary}) to hold. While one can speculate that this discrepancy may arise from not carefully taking into account the change in the functional form of $C(t)$ below $k_\dagger$ (cf.~Fig.~\ref{fig:Fig2}c) when extracting attenuation rates, we presently cannot fully explain this striking disagreement with our theoretical arguments and numerical results.

\section{The generalized-Rayleigh scaling originates from quasilocalized nonphononic excitations}
\label{sec:log}

In previous sections we provided substantial evidence that the abundance of quasilocalized modes at frequencies in the vicinity of and above the crossover frequency $\omega_\dagger$ gives rise to the generalized-Rayleigh scaling of wave attenuation rates. In order to further cement this conclusion, we seek to study model computer glasses in which the abundance of quasilocalized modes at low frequencies is controllable by tuning an external model parameter. By the controlled depletion of soft quasilocalized modes, we expect to be able to systematically shift the crossover wavenumber $k_{\rm on}$ that separates the generalized-Rayleigh scaling of attenuation rates to Rayleigh scattering scaling.

\begin{figure}[ht!]
  \centering
  \includegraphics[width=\columnwidth]{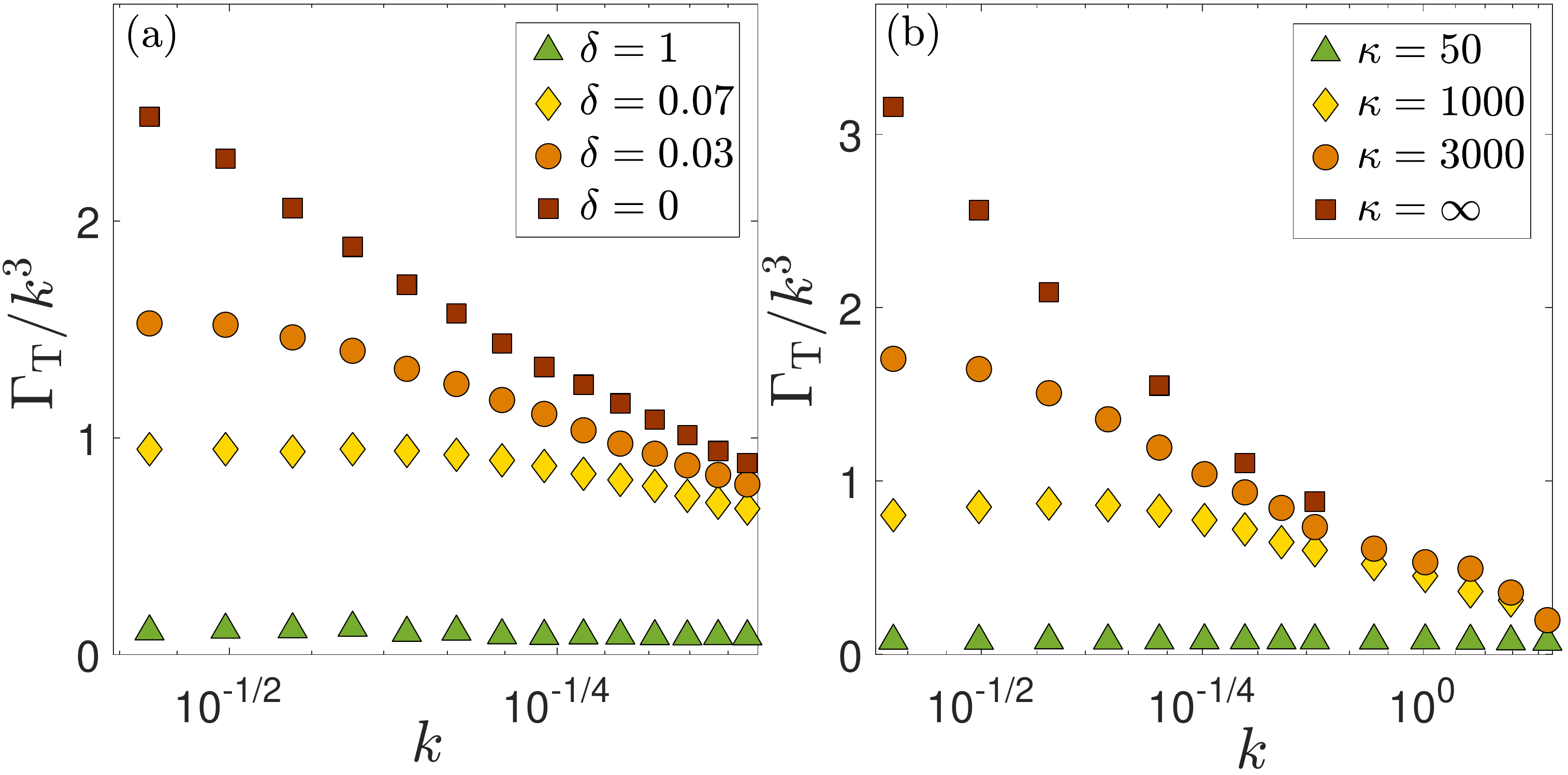}
  \caption{(a) Rescaled transverse (shear) wave attenuation rates $\Gamma_{\rm T}(k;\delta)/k^3$ measured for dynamics governed by the Hessian as given by Eq.~(\ref{eq:augmented_hessian}), for various values of the parameter $\delta$, and $N\!=\!25600$. As predicted, increasing $\delta$ shifts the crossover wavenumber $k_{\rm on}(\delta)$, from Rayleigh to generalized-Rayleigh scaling, to higher values. (b) The same as (a), but measured in computer glasses of $N\!=\!1$M particles created using a model that allows particle sizes to fluctuate during glass formation (but not afterwards, see~\cite{fsp_pre} and text for details). Particle-size fluctuations during glass formation are governed by the parameter $\kappa$, with $\kappa\!=\!\infty$ meaning no fluctuations. Similar to (a), here we find that $k_{\rm on}(\kappa)$ shifts to higher wavenumbers when $\kappa$ is decreased, as predicted.}\label{fig:Fig5}
\end{figure}

To this aim we first consider glassy samples of the 2DIPL model; in simple models of computer glasses such as the 2DIPL model, the Hessian matrix can be decomposed as $\calBold{M}\!=\!\calBold{M}_\kappa + \calBold{M}_f$, where $\calBold{M}_\kappa$ accounts for pairwise stiffnesses, and $\calBold{M}_f$ accounts for pairwise internal-stresses/forces~\cite{eric_boson_peak_emt, inst_note}. To control the abundance of quasilocalized glassy modes, we employed an augmented Hessian that reads
\begin{equation}
\label{eq:augmented_hessian}
\calBold{M}(\delta) = \calBold{M}_\kappa + (1-\delta)\calBold{M}_f\,,
\end{equation}
where $\delta$ is a dimensionless control parameter. In previous work~\cite{eric_boson_peak_emt,inst_note} it was argued that increasing $\delta$ from zero creates a gap in the nonphononic density of states $D_{\rm G}(\omega)$ at zero frequency, which grows at small $\delta$ as $\sim\!\sqrt{\delta}$. We thus expect wave attenuation rates measured in a system governed by a Hessian of the form of Eq.~(\ref{eq:augmented_hessian}) to feature a crossover from Rayleigh scattering scaling to generalized-Rayleigh scaling at $k_{\rm on}(\delta)$, which is now expected to increase as $\delta$ is increased. This expectation is indeed verified in Fig.~\ref{fig:Fig5}a.

We next consider the 2D variant of the model computer glass introduced in~\cite{fsp_pre}, referred to as the 2D Fluctuating-Size Particles (2DFSP) model. The 2DFSP model consists of particles whose sizes are considered as degrees of freedom that are subjected to a potential of the form of a well, characterized by a stiffness $\kappa$. Within the framework of this model, particle sizes are allowed to fluctuate during glass formation, but, upon completion of glass formation, particle sizes are fixed for all subsequent analyses. In~\cite{fsp_pre} it was shown that tuning $\kappa$ (denoted by $k_\lambda$ there) has a dramatic effect on the elastoplastic properties of the resulting glasses. In particular, it was shown that the nonphononic density of states features a gap at zero frequency that scales as $1/\sqrt{\kappa}$ at large $\kappa$. We therefore expect wave attenuation rates to show a crossover from Rayleigh scattering scaling to generalized-Rayleigh scaling at a crossover wavenumber $k_{\rm on}(\kappa)$ that should increase with decreasing $\kappa$. This expectation is indeed verified in Fig.~\ref{fig:Fig5}b.

The nature of the nonphononic DOS of the two models investigated in this section is anomalous: generic glass formers, and even very well-annealed model glasses~\cite{LB_modes_2019}, feature the universal $D_{\rm G}(\omega)\!\sim\!\omega^4$. Notwithstanding, by exploiting the control offered by computer models, we are able to firmly establish that soft quasilocalized modes are the origin of the generalized-Rayleigh scaling of wave attenuation rates, and not correlations in local elastic moduli fields (which are present for all $\kappa$ values in the FSP model~\cite{fsp_pre}) as suggested in~\cite{lemaitre_tanaka_2016}.

\section{Experimental evidence}
\label{sec:exp}

\begin{figure}[ht!]
  \centering
  \includegraphics[width=0.93\columnwidth]{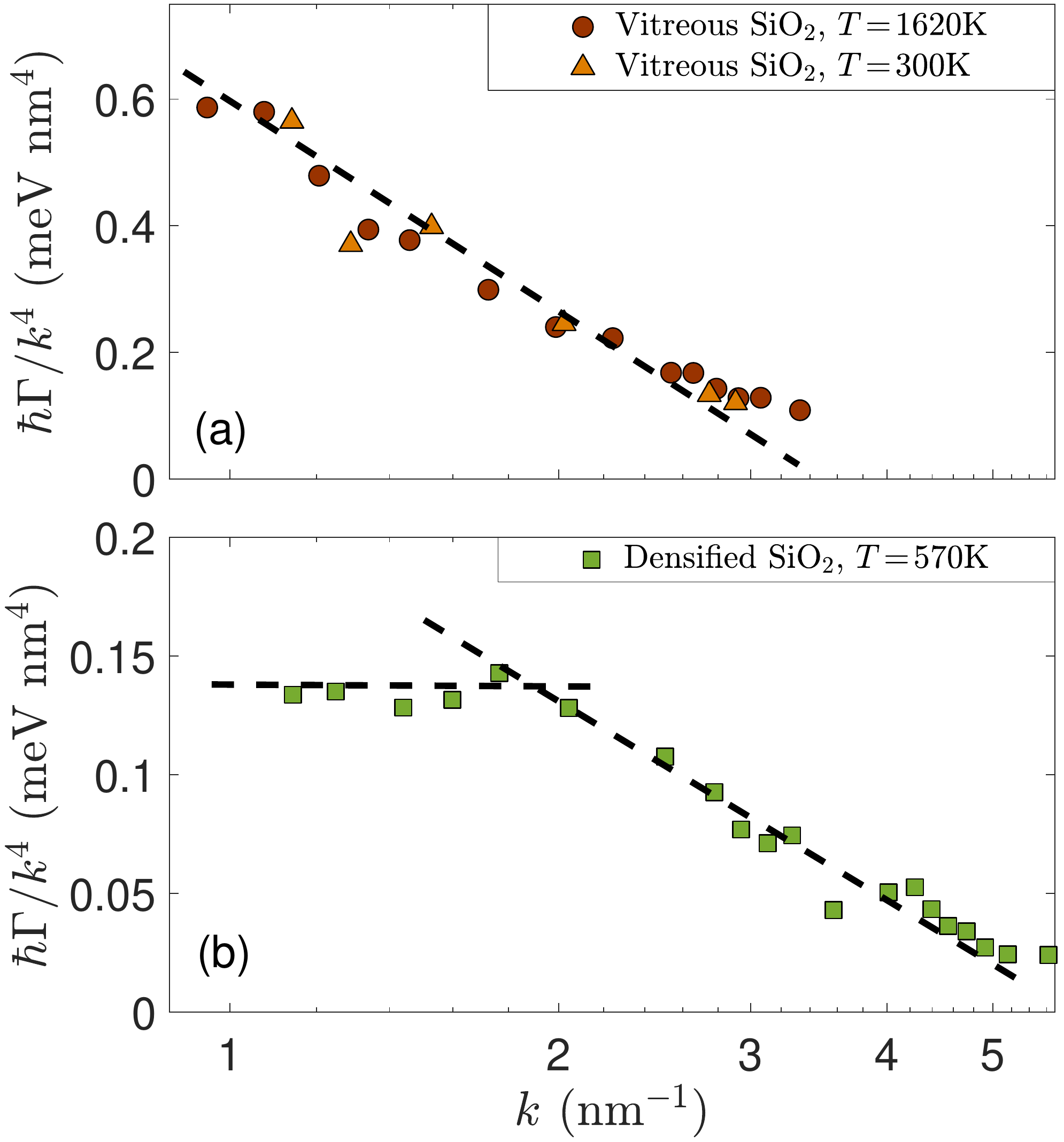}
  \caption{Wave attenuation rates, rescaled by $k^4$, measured in experiments on vitreous (panel~(a), data digitized from~\cite{experiments_300K_vSiO2,experiments_1620K_vSiO2}) and densified (panel~(b), data digitized from~\cite{experiments_570K_dSiO2}) SiO$_2$, using the high-resolution inelastic x-ray scattering technique, see text for discussion. Panel (b) exhibits a crossover from generalized-Rayleigh to Rayleigh scaling (dashed lines), as predicted.}\label{fig:Fig6}
\end{figure}

As recently pointed out~\cite{lemaitre_tanaka_2016}, several experimental data sets support that wave attenuation rates in laboratory glasses follow the generalized-Rayleigh scaling over some wavenumber range. In Fig.~\ref{fig:Fig6}a we present digitized data from~\cite{experiments_300K_vSiO2,experiments_1620K_vSiO2} of wave attenuation rates in vitreous silica (silicon dioxide, SiO$_2$) measured down to $k\!\simeq\!1$nm$^{-1}$ using high-resolution inelastic x-ray scattering (IXS)~\cite{Ruocco1996, ixs_review_arXiv}. Note that the experimental data correspond to two different temperatures (see legend), with no noticeable temperature dependence, indicating the absence of significant anharmonic effects (additional evidence for the existence of a sizable temperature range that is dominated by harmonic effects is presented in~\cite{Ruffle2006,Ruta2010}). Consequently, our predictions in the harmonic regime are expected to be directly relevant for these experimental data. While these data --- presented and discussed also in~\cite{lemaitre_tanaka_2016} --- are consistent with generalized-Rayleigh scaling, the results extensively discussed above predict that a crossover to Rayleigh scaling should be observed at yet smaller wavenumbers $k$. However, present-day IXS spectrometers operate at momenta transfer corresponding to $k\!>\!1$nm$^{-1}$, so this lower $k$ regime is currently inaccessible using this experimental technique. Yet, as shown above, suppressing the density of soft quasilocalized modes pushes the crossover to Rayleigh scaling to higher wavenumbers, which in principle could make the crossover observable using present-day IXS. This might be achieved using a permanent densification procedure~\cite{densification}.

Permanent densification of glassy materials has been previously shown to greatly suppress the magnitude of the Boson peak in the DOS of glasses~\cite{Effect_of_Densification_prl_2006}. We hypothesize that permanent densification likewise leads to the annihilation and stiffening of soft quasilocalized modes, thus suppressing their DOS at low wavenumbers. In turn, this densification process potentially pushes the crossover wavenumber $k_{\rm on}$ into the wavenumber regime accessible to IXS. To test this idea, we present in Fig.~\ref{fig:Fig6}b digitized data from~\cite{experiments_570K_dSiO2} of wave attenuation rates measured (by the same experimental group that measured the data presented in Fig.~\ref{fig:Fig6}a) in silica that was subjected to permanent densification under a pressure of $8$GPa (see~\cite{experiments_570K_dSiO2} for the details of the densification procedure). Strikingly, a clear crossover from generalized-Rayleigh to Rayleigh scaling at $k_{\rm on}\!\approx\!2$nm$^{-1}$ is observed, strongly supporting the physical picture discussed in this paper. We note that according to a recent study~\cite{Deschamps2014}, densification of vitreous silica at $8$GPa does not lead to significant changes in the elastic moduli. Consequently, the differences observed in attenuation rates between vitreous and densified silica could not be attributed to shifts in characteristic phononic frequencies; rather, we propose that they emerge from the suppression of soft nonphononic quasilocalized modes.

The experimental data set shown in Fig.~\ref{fig:Fig6}b supports our assertion that attenuation rates in macroscopic glasses should feature Rayleigh scaling at the lowest wavenumbers $k$ and a crossover to generalized-Rayleigh scaling at higher $k$. Clearly, additional measurements of attenuation rates at sufficiently lower wavenumbers in glasses are needed to further test our predictions. Moreover, improved control over structural disorder in glasses, e.g.~as recently done in~\cite{Ketkaew2018}, can affect the crossover to Rayleigh scaling and make it accessible to present-day measurement techniques.

\section{Concluding remarks and open questions}
\label{sec:sum}

In this paper, we have theoretically and computationally established three basic results in glass physics. First, we have shown that the finite-size quantization of phonons into discrete bands in finite glassy samples gives rise to qualitative changes in the physics of wave attenuation: for wavenumbers $k$ below the $N$-dependent crossover wavenumber $k_\dagger(N)$, attenuation rates follow the finite-size scaling theory Eq.~(\ref{eq:scattering_summary}), meaning that they depend explicitly on system size, and on the number of phononic modes of the corresponding discrete phonon band. Above the crossover $k_\dagger$, attenuation rates become independent of system size, and only depend on the wavenumber. In addition, the functional form of the velocity autocorrelation function, from which the attenuation rates are extracted, changes qualitatively between above and below the crossover.

The second key result we have established is that the recently-observed generalized-Rayleigh scaling of attenuation rates originates from the presence and number of soft quasilocalized modes emerging from glassy disorder/microstructures, whose universal statistical and structural properties have recently been revealed~\cite{modes_prl_2016, modes_prl_2018, ikeda_pnas}. The generalized-Rayleigh regime corresponds thus to a distinct frequency regime in the vibrational DOS --- below the boson peak frequency --- in which phonons and an abundance of soft quasilocalized modes coexist (in hybridized form, see discussions in e.g.~\cite{SciPost2016,phonon_widths}). It therefore constitutes a distinct wave attenuation regime, and not merely a crossover between the Rayleigh ($\sim\!k^4$) and the `boson peak' ($\sim\!k^2$) regimes.

Finally, we have shown that at very low wavenumbers corresponding to frequencies at which soft quasilocalized modes are scarce, wave attenuation rates follow Rayleigh scattering scaling. This implies that in macroscopic glasses Rayleigh scaling is the expected asymptotic low-wavenumber behavior of wave attenuation rates, and not generalized-Rayleigh scaling as recently claimed~\cite{lemaitre_tanaka_2016}. Generalized-Rayleigh scaling is expected to be observed at higher wavenumbers. These predictions are further supported by experimental data from recent literature~\cite{experiments_300K_vSiO2,experiments_1620K_vSiO2,experiments_570K_dSiO2}.

These results raise various interesting questions, both theoretical and experimental; among these, we would like to highlight the most prominent and pressing ones, in our view. First, we need to understand how quasilocalized nonphononic modes give rise to the generalized-Rayleigh scaling of attenuation rates once they are sufficiently abundant. That is, we need to develop a first principles theory of the apparently universal scaling relation $\Gamma(k)\!\sim\!k^{\dbar+1}\log\!{(k_0/k)}$ for $k\!>\!k_{\rm on}$ and of the crossover wavenumber $k_{\rm on}$, based on the physical insight gained in this work.

Second, we need to explore the implications of the obtained results for $\Gamma(k)$ at small wavenumbers $k$ for the low temperature energy diffusivity and heat transport in glasses~\cite{Kittel_1949,Zeller_and_Pohl_prb_1971}. In addition, we need improved experiments, guided by theoretical estimates of $k_{\rm on}$, which aim at directly measuring the crossover from Rayleigh to generalized-Rayleigh scaling of long wavelength wave attenuation rates in glasses. We hope that these important challenges will be addressed in future work.

Finally, in a broader context, it would be interesting to understand whether the ideas developed in this paper might be related to universal low-temperature thermodynamic and transport anomalies in glasses. It is well-established that the specific heat $C(T)$ of glasses scales linearly with the temperature $T$ below $1$K, instead of Debye's $T^3$ prediction for crystals~\cite{Zeller_and_Pohl_prb_1971}, and exhibits a distinct hump around $10$K, when normalized by $T^3$~\cite{soft_potential_model_1991}. In the very same temperature range, $\sim\!1\!-\!10$K, the thermal conductivity $\kappa(T)$ of glasses exhibits a plateau that is entirely absent in their crystalline counterparts~\cite{Eucken_1911,Zeller_and_Pohl_prb_1971,Freeman_Anderson_prb_1986}. Despite extensive efforts~\cite{Kittel_1949,Anderson,Phillips,Tanguy_pre_2018,moshe_prb_2013,Schober_prb_1992,energy_transport_jamming,Gurevich2003,ganter2010rayleigh,schirmacher2011comments,Maurer2004,mw_EM_epl,eric_boson_peak_emt}, a complete and fundamental understanding of these universal low-temperature glass anomalies is currently missing. Future work should clarify whether the physics behind sound attenuation discussed in this paper might be related to these universal low-temperature anomalies.

\textit{Acknowledgments.--} E.~L.~acknowledges support from the Netherlands Organisation for Scientific Research (NWO) (Vidi grant no.~680-47-554/3259). E.~B.~acknowledges support from the Minerva Foundation with funding from the Federal German Ministry for Education and Research, the William Z.~and Eda Bess Novick Young Scientist Fund and the Harold Perlman Family.

\appendix

\section{numerical models and methods}
\label{sec:numerics_appendix}

In this Appendix we provide details about the computer glass forming models employed, and about the methods used to measure wave attenuation rates in our glassy samples. We reiterate that all observables reported in the main text are made dimensionless as follows: lengths are reported in terms of $a_0\!\equiv\! (V/N)^{1/\dbar}$ where $V,N,\dbar$ denote the volume, number of particles, and spatial dimension, respectively; times (rates) are reported in units of $a_0/c_{\rm T}$ ($c_{\rm T}/a_0$) where $c_{\rm T}\!\equiv\!\sqrt{G/\rho}$ denotes the shear wave speed, $\rho\!=\! mN/V$ denotes the mass density, $m$ is the particles' mass, and $G$ is the athermal shear modulus, defined as \cite{lutsko}
\begin{equation}
G \equiv \frac{1}{V}\left(\frac{\partial^2U}{\partial\gamma^2}  - \frac{\partial^2U}{\partial\gamma\partial\rv_i}\cdot\calBold{M}_{ij}\cdot\frac{\partial^2U}{\partial\rv_j\partial\gamma}\right)\,,
\end{equation}
where $U$ is the potential energy, $\rv_i$ denotes the $i^{\mbox{\tiny th}}$ particle's coordinates, $\calBold{M}_{ij}\!\equiv\!\frac{\partial^2U}{\partial\rv_i\partial\rv_j}$ is the Hessian matrix, and $\gamma$ is a shear strain deformation.

\subsection{Computer glass forming models}
In this work we employ three different computer glass forming models in 2D and 3D.

\subsubsection{Inverse power law (IPL)}
The inverse power law (IPL) model is a 50:50 binary mixture of `large' and `small' particles that interact via a $r^{-10}$ purely repulsive pairwise potential. Details about the model can be found in \cite{cge_paper,modes_prl_2018}, and the preparation protocol we have followed is described in \cite{modes_prl_2018}. We employ the same model in 2D and 3D (however with different densities, see \cite{modes_prl_2018}), which are referred to in the main text as 2DIPL and 3DIPL, respectively. Glassy samples of these models were prepared by a continuous quench from  high temperature equilibrium states, at the highest rate that produces glasses that feature the universal nonphononic DOS $D_{\rm G}(\omega) \sim \omega^4$, as shown in \cite{protocol_prerc}. As a result, these glasses feature an abundance of soft quasilocalized modes at low frequencies, as shown in \cite{modes_prl_2018, phonon_widths}. Attenuation rates were measured in at least 125 independent glassy samples of each system size of the 2DIPL model, and at least 100 independent glassy samples for each system size of the 3DIPL model.

\subsubsection{Polydisperse swap Monte-Carlo (SWAP)}
We employed a slightly modified version of the polydisperse computer glass forming model introduced in \cite{berthier_prx}. This model was shown to be robust against crystallization, and at the same time allows for extremely deep supercooling by means of the swap Monte-Carlo algorithm \cite{TSAI1978465, GAZZILLO1989388, Grigera2001}. Details about the preparation protocol of glassy samples, and specific details about the interaction potential can be found in~\cite{boring_arXiv_2019}. We emphasize that sample-to-sample fluctuations in elastic properties, which arise due to the broad distribution from which particle sizes are drawn, are handled as described in~\cite{boring_arXiv_2019}.

We chose the densities $\rho\!=\!0.65$ and $\rho\!=\!0.58$ (expressed in the units provided in \cite{boring_arXiv_2019}) in 2D and 3D, respectively. The equilibrium temperatures from which our glassy samples were instantaneously quenched were approximately 0.45$T_{\rm x}$ in 3D and 0.25$T_{\rm x}$ in 2D, where $T_{\rm x}$ denotes the temperature below which the shear modulus of underlying inherent states starts to depend on the parent temperature from which glassy samples were instantaneously quenched \cite{boring_arXiv_2019}. Attenuation rates were measured in $1500$ independent glassy samples of $N\!=\!16384$ particles of the 2DSWAP model, and in at least 50 independent glassy samples of each system size of the 3DSWAP model.

\subsubsection{2D Fluctuating-Size Particles (2DFSP)}
Glasses created with the Fluctuating-Size Particles protocol are made by instantaneously quenching a high-temperature liquid in an augmented potential energy landscape, that includes the particle radii as additional degrees of freedom~\cite{fsp_pre}. The potential that governs the size degrees of freedom is characterized by a stiffness $\kappa$; setting $\kappa\!=\!\infty$ corresponds to freezing out the size degrees of freedom completely, i.e.~a regular quench. In all subsequent analyses of the resulting polydisperse glasses, particle radii are frozen, i.e.~no longer considered degrees of freedom. The samples used in this work were made with the model and protocol described in \cite{fsp_pre}, with the only difference being that we worked in 2D instead of 3D. We generated 100 independent glassy samples of $N\!=\!10^6$ particles for each value of $\kappa$.

\subsection{Measurement of wave attenuation rates}
\label{sec:fitting_procedure}
We measure wave attenuation rates in our computer glasses following the procedure presented in~\cite{lemaitre_tanaka_2016}: given a glassy sample in mechanical equilibrium (i.e.~in a minimum of the potential energy $U$), we choose a wavevector ${\bm k}$ and impose a velocity field of the form given by Eq.~(\ref{eq:phonon}). Newtonian dynamics are then integrated forward in time for $t\!>\!0$ within the harmonic approximation, namely
\begin{equation}
m\ddot{\uv} = -\calBold{M}\cdot\uv\,,
\end{equation}
where $\uv$ denotes the displacement of particles' positions about the potential energy minimum. The velocity autocorrelation function $C(t)$ as given by Eq.~(\ref{eq:corr}) is monitored for each such run, and averaged over many runs (between $200$ and several thousands) executed for independent glassy samples and different wavevectors ${\bm k}$ that share the same wavenumber $k\!=\!|{\bm k}|$.

Wave attenuation rates $\Gamma_{\rm T,L}(k)$ are obtained by extracting the envelopes of the averaged velocity autocorrelation functions. For wavenumbers $k\!<\!k_\dagger$ in the finite-size regime (see main text for detailed discussions) we fit the envelopes to a compressed exponential form $C(t)\!=\!\exp\big[-\frac{1}{2}(\Gamma t)^\alpha\big]$ with $\Gamma$ and $\alpha$ serving as fitting parameters. For wavenumbers $k\!>\! k_\dagger$ above the finite size regime, the envelopes of $C(t)$ follow an exponential form (see Fig.~\ref{fig:Fig1}b,c), and we therefore fit the velocity autocorrelation functions to an exponential decay $C(t)\!=\!\exp\big[-\frac{1}{2}\Gamma t\big]$, where $\Gamma$ is the only fitting parameter.
\begin{figure}[ht!]
  \centering
  \includegraphics[width=0.95\columnwidth]{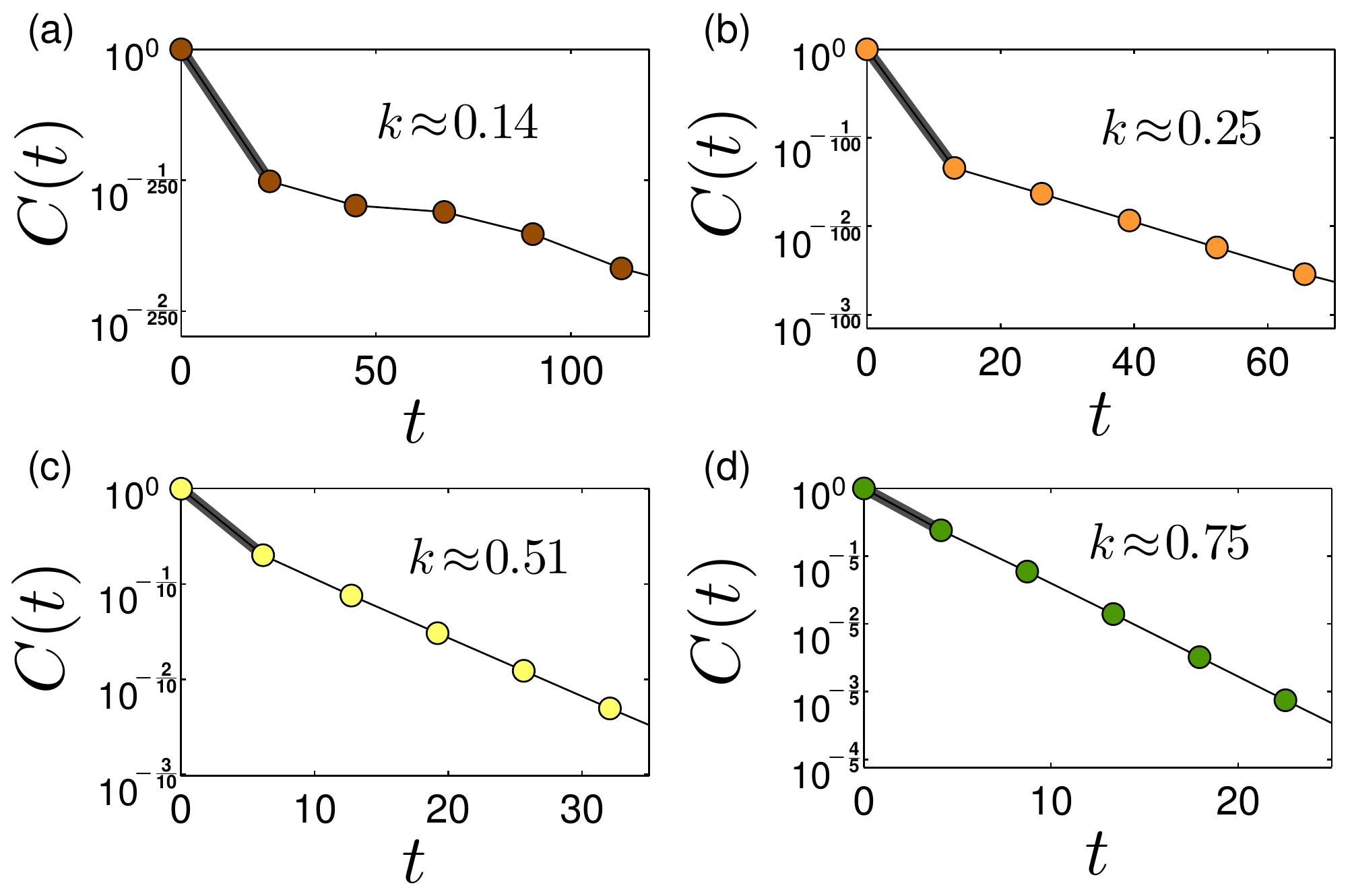}
  \caption{Panels (a)-(d) show the envelopes of the velocity correlation functions $C(t)$ during the first 5 oscillations, measured for the 3DSWAP model with $N\!=\!256$K particles for different wavenumbers as indicated by the legends. We observe an initial drop in the first oscillation --- marked by the thick gray lines --- of magnitude $\sim\! k^2$, both in 2D and 3D.}\label{fig:Fig7}
\end{figure}

We note that the envelopes of the velocity correlation functions $C(t)$ exhibit an initial drop at the first oscillation (i.e.~when the first minimum occurs), as shown in Fig.~\ref{fig:Fig7}. Plotted are the envelopes of the correlation functions during the first $5$ oscillations, for different wavenumbers $k$ as seen in the legends. The initial drops/decays, marked by thick gray lines, are found to scale as $k^2$, both in two and three dimensions. Finally, it should be noted that while the initial decays increase with increasing $k$, only at small wavenumvers $k$ a discontinuity in the envelope derivative $dC(t)/dt$ exists.

\section{$C(t)$ in the finite-size regime}
\label{sec:networks_appendix}
In \cite{Grzegorz_scattering_arXiv2} it was claimed that the envelopes of the velocity correlation functions $C(t)$ measured for $k\!<\! k_\dagger$ in stable computer glasses feature an initial exponential decay over an $N$-dependent range, rather than a compressed-exponential decay as we reported in Fig.~\ref{fig:Fig1} and motivated in Sect.~\ref{sec:2dipl}. In order to shed light on this issue, we seek to employ a system which possesses no quasilocalized modes in between its discrete phonon bands, and as such features very well-isolated phonon bands. As we argued in Sect.~\ref{sec:2dipl}, this set up should give rise to compressed-exponentially decaying velocity correlation functions $C(t)$.

Random networks of relaxed Hookean springs are a suitable choice to this aim, as they feature no quasilocalized modes \cite{inst_note,ikeda_pnas}. We created random spring networks from our 3DIPL glassy samples by placing a relaxed (i.e.~at its rest-length) Hookean spring with unit stiffness between every pair of interacting particles of the original glass. We then measured the velocity correlation functions $C(t)$ as explained in Sect.~\ref{sec:theory}, for wavenumbers that correspond to the first $6$ lowest frequency phonon bands.
\begin{figure}[ht!]
  \centering
  \includegraphics[width=0.95\columnwidth]{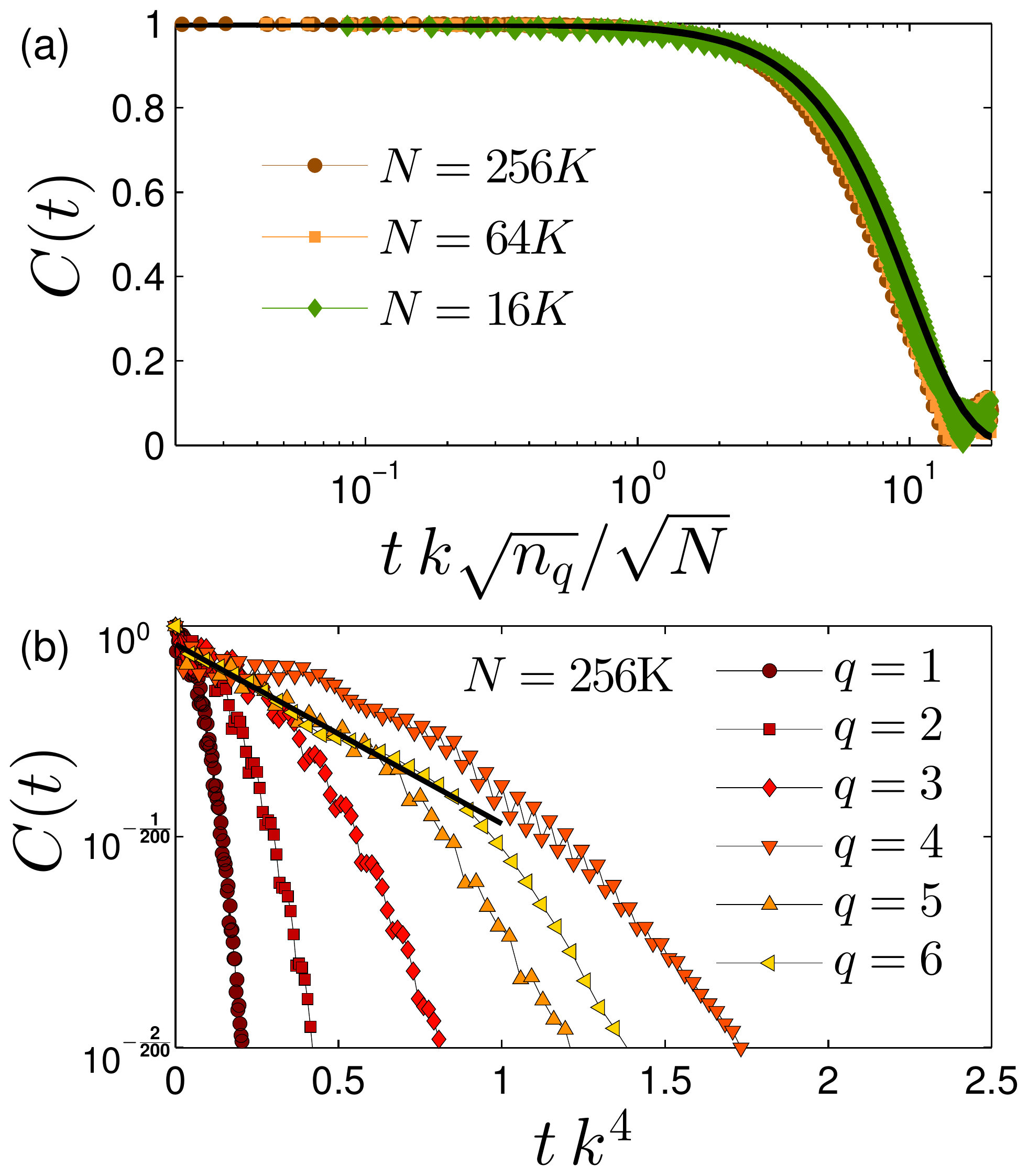}
  \caption{Envelopes of the velocity correlation functions $C(t)$ as defined in Eq.~(\ref{eq:corr}), measured in 3D disordered networks of Hookean springs for the first 6 lowest $q$'s, plotted against $tk\sqrt{n_q}/\sqrt{N}$ in panel (a), and against $tk^4$ in panel (b). The black lines correspond to Gaussian (panel (a)) and exponential (panel (b)) decays, see text for further details.}\label{fig:Fig8}
\end{figure}

The results of these measurements are plotted in Fig.~\ref{fig:Fig8}; in panel (a) we plot the correlation functions against the time rescaled by the predicted width of the corresponding phonon band $\Delta\omega\!\sim\! k\sqrt{n_q}/\sqrt{N}$, which results in an excellent collapse of the correlation functions over the entire time domain, similar to the results presented in \cite{phonon_widths} for disordered lattices, and validating once again our finite-size scaling theory for wave attenuation rates. The continuous black line --- that represents a Gaussian fit --- captures very well the precise functional form of the correlation functions across the majority of the time axis.

In Fig.~\ref{fig:Fig8}b we show the very initial stages of the decay of $C(t)$ for the largest systems simulated, plotted against the rescaled time $tk^4$ (notice the logarithmic scale of the $y$-axis). If Rayleigh scattering can be seen at short times, as claimed in \cite{Grzegorz_scattering_arXiv2}, the signals should collapse and feature an initial exponential functional form. We see that a first indication of an initial exponential decay (continuous black line) can be seen in for $q\!=\!5$ and $q\!=\!6$, albeit over a very short time range, in agreement with the claims of \cite{Grzegorz_scattering_arXiv2}. The lower-$q$ bands, however, do not collapse, nor do they feature an exponential decay, indicating that the method suggested in \cite{Grzegorz_scattering_arXiv2} to extract macroscopic wave attenuation rates in the finite-size regime is limited -- its validity might not extend down to the lowest accessible wavenumbers.

\begin{figure}[ht!]
  \centering
  \includegraphics[width=\columnwidth]{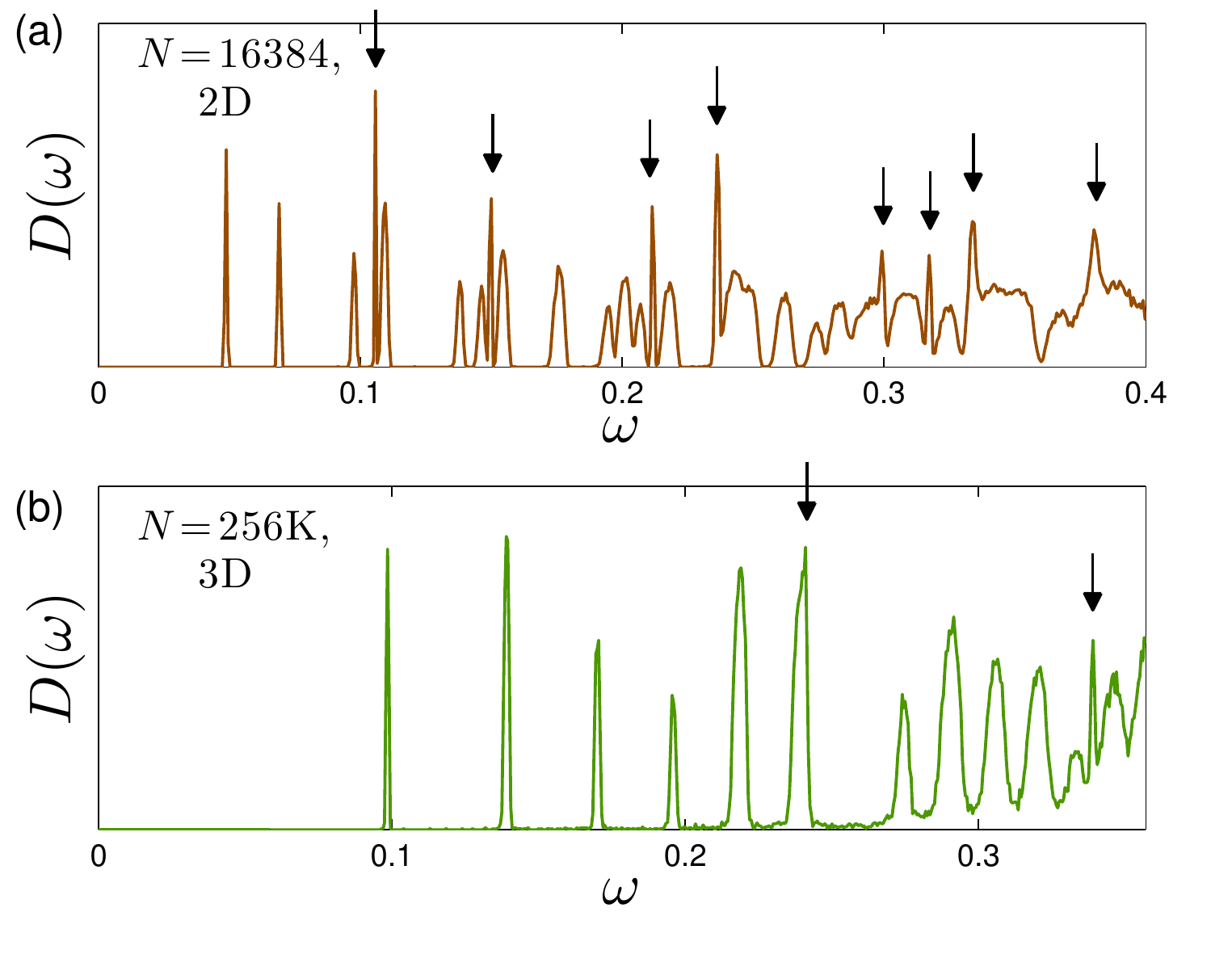}
  \caption{Vibrational density of states $D(\omega)$ measured in (a) the 2DSWAP and (b) the 3DSWAP models. The arrows indicate the positions of sound (longitudinal) waves, see text for discussion.}\label{fig:Fig9}
\end{figure}

\section{Attenuation of long-wavelength sound (longitudinal) waves}
\label{sec:soundwaves_appendix}

In Fig.~\ref{fig:Fig9} we show the vibrational density of states of stable glasses in 2D and 3D, of sizes as indicated by the legends. In our 2D system we find several sound waves below the crossover frequency $\omega_\dagger$ above which the widths of phonon bands becomes comparable to the gaps between adjacent bands. In 3D, for the system size analyzed $(N\!=\!256$K) only a single sound (longitudinal) wave resides below $\omega_\dagger$, and is itself hybridized with the 6$^{\mbox{\tiny th}}$ shear wave band. We assert that since all but a single sound wave reside above $\omega_\dagger$ in stable glasses of $\approx\!10^5$ particles in 3D, the attenuation rates pertaining to those sound waves would show no finite-size effects, explaining the observation reported in~\cite{Grzegorz_scattering_arXiv}.

\begin{figure}[ht!]
  \centering
  \includegraphics[width=0.82\columnwidth]{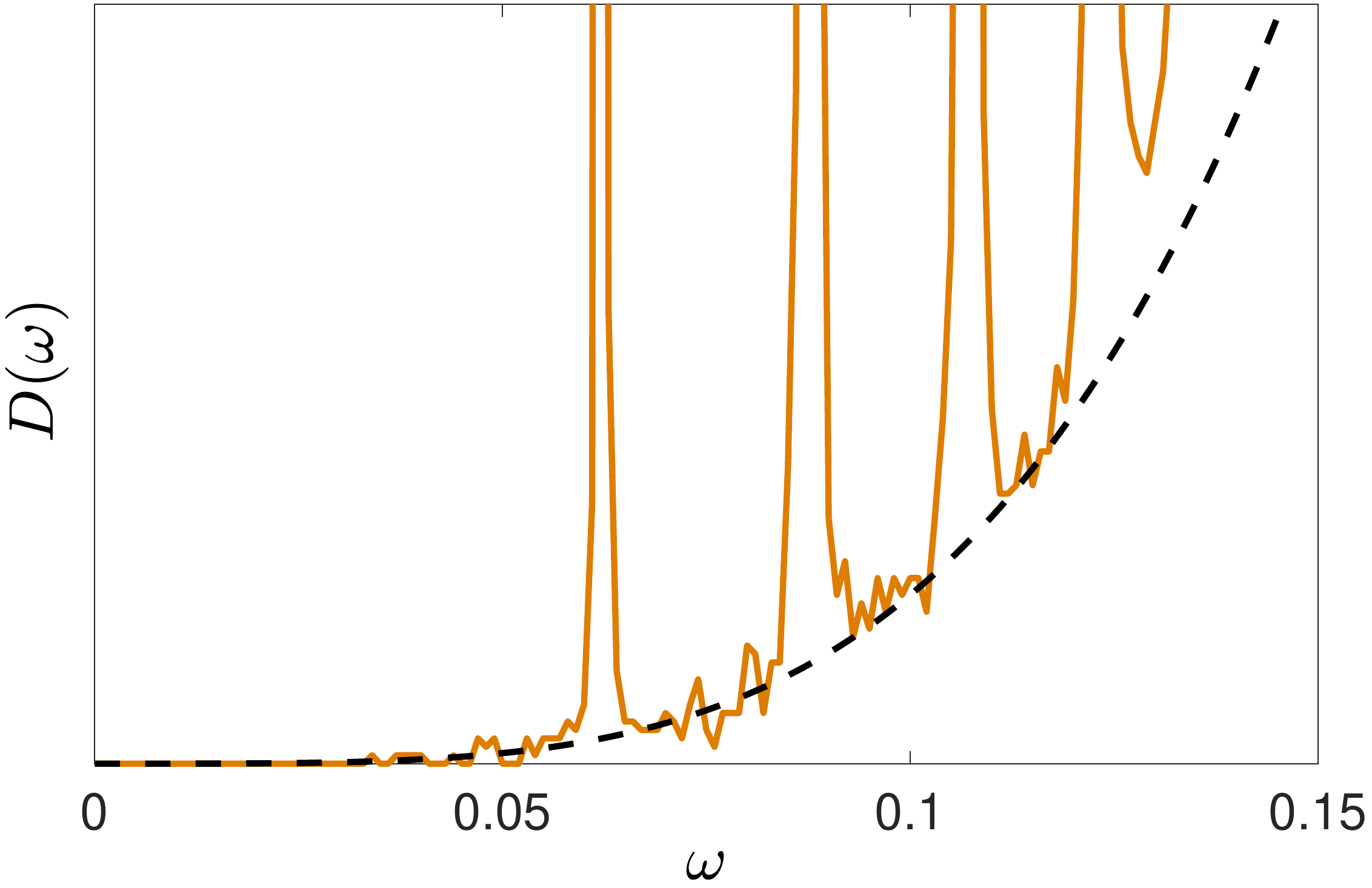}
  \caption{Vibrational density of states $D(\omega)$ measured in 3DIPL glasses with $N\!=\!1$M, corresponding to one of the data sets presented in Fig.~\ref{fig:Fig4}. The dashed line, which indicates the lower envelope of $D(\omega)$, corresponds to the density of states of nonphononic quasilocalized modes, $D_{\rm G}(\omega)\!\sim\!\omega^4$. It is clear that $k_{\rm on}$ resides well-within the quasilocalized modes frequency regime (recall that in our dimensionless units $\omega\!\approx\! k$).}\label{fig:Fig10}
\end{figure}

\section{The nonphononic density of states}
\label{sec:DOS_appendix}

In Fig.~\ref{fig:Fig10} we show the vibrational density of states of 3DIPL glasses with $N\=1$M, corresponding to one of the data sets presented in Fig.~\ref{fig:Fig4}. It is observed that while the phononic part of the DOS shown in the figure is clearly dominated by finite-size effects, manifested by the discrete (and finite width) bands, the lower envelope of the DOS follows a continuous contribution that corresponds to the nonphononic part of the DOS, $D_{\rm G}(\omega)\!\sim\!\omega^4$ (dashed line).

This result shows that the transition from Rayleigh to generalized-Rayleigh scaling regimes, observed in Fig.~\ref{fig:Fig4}b to occur for significantly larger frequencies/wavenumbers (beyond the finite-size regime), occurs well within the regime in which $D_{\rm G}(\omega)\!\sim\!\omega^4$ is realized in our calculations.


\begin{thebibliography}{61}
\bibitem [{foo({\natexlab{b}})}]{footnote}%
  \BibitemOpen
  \href@noop {} {\bibfield  {journal} {\bibinfo  {journal} {This is done in
  models with pairwise radially-symmetric potentials by considering the
  truncated dynamical matrix
  $\tilde{\calBold{M}}\!\equiv\!\sum_{i<j}\varphi''_{ij}\nv_{ij}\nv_{ij}$ with
  $\varphi_{ij}$ and $\nv_{ij}$ the pairwise potential and director,
  respectively}\ } ({\natexlab{b}})}\BibitemShut {NoStop}%
\end{thebibliography}
%

\end{document}